\documentclass[10pt,twocolumn]{article}
\usepackage{pslatex}
\usepackage[T1]{fontenc}
\usepackage{epsfig}
\usepackage{setspace}
\usepackage{subfigure}
\usepackage[left=1in,top=1in,right=1in,bottom=1in,nohead,foot=0.5in]{geometry}
\usepackage[small]{caption}
\long\def\symbolfootnote[#1]#2{\begingroup%
\def\thefootnote{\fnsymbol{footnote}}\footnote[#1]{#2}\endgroup} 

\title{\vspace{-0.5in}Running a Quantum Circuit at the Speed of Data}
\author{Nemanja Isailovic, Mark Whitney, Yatish Patel and John Kubiatowicz \\
Computer Science Division \\
University of California, Berkeley \\
\{nemanja, whitney, yatish, kubitron\}@cs.berkeley.edu \\[-0.2in]
} % end author

\date{}%{\today at \ampmtime }

\begin{document}
\maketitle
%\vspace{-0.9in}
\symbolfootnote[0]{To appear in the 35th annual International Symposium on Computer Architecture (ISCA '08)}
\newcommand{\tOneQGate}{1}
\newcommand{\tTwoQGate}{10}
\newcommand{\tMeas}{50}
\newcommand{\tPhysPrep}{51}
\newcommand{\tMoveLat}{1}
\newcommand{\tTurnLat}{10}
\newcommand{\verifSuccessRate}{0.998}
\newcommand{\qecStageTwo}{20}
\newcommand{\qecStageThree}{51}
\newcommand{\steaneLOneEncZeroPrepLat}{}
\newcommand{\steaneLOneVerify}{}
\newcommand{\steaneLOneGoodEncZeroPrepLat}{213}
\newcommand{\steaneLOneGoodQECLat}{284}
\newcommand{\smallCircuitLat}{22}
\newcommand{\smallCircuitLatWQEC}{590}
\newcommand{\steaneLTwoStageOneTime}{568}
\newcommand{\steaneLTwoTotalTime}{639}
\newcommand{\golayTotalTime}{0}
\newcommand{\golayStageOneTime}{0}
\newcommand{\nonOverlappedQECLat}{517}
\newcommand{\simpleCircuitLatWithoutQEC}{22}
\newcommand{\simpleCircuitLatWithQEC}{2090}
\newcommand{\qrcaNumQubits}{97}
\newcommand{\qclaNumQubits}{123}
\newcommand{\qftNumQubits}{32}
\newcommand{\qrcaTwoQGateCount}{884}
\newcommand{\qrcaOneQTransGateCount}{378}
\newcommand{\qrcaOneQNonTransGateCount}{882}
\newcommand{\qrcaPrepareGateCount}{33}
\newcommand{\qrcaTotalGateCount}{2177}
\newcommand{\qrcaOneQNonTransGatePercent}{40.5}
\newcommand{\qclaTwoQGateCount}{941}
\newcommand{\qclaOneQTransGateCount}{423}
\newcommand{\qclaOneQNonTransGateCount}{987}
\newcommand{\qclaPrepareGateCount}{59}
\newcommand{\qclaTotalGateCount}{2410}
\newcommand{\qclaOneQNonTransGatePercent}{41.0}
\newcommand{\qftTwoQGateCount}{145}
\newcommand{\qftOneQTransGateCount}{4184}
\newcommand{\qftOneQNonTransGateCount}{3816}
\newcommand{\qftPrepareGateCount}{0}
\newcommand{\qftTotalGateCount}{8145}
\newcommand{\qftOneQNonTransGatePercent}{46.9}
\newcommand{\qrcaCritPath}{29508}
\newcommand{\qclaCritPath}{3827}
\newcommand{\qftCritPath}{77057}
\newcommand{\qrcaQECLatAncPrep}{447726}
\newcommand{\qclaQECLatAncPrep}{55806}
\newcommand{\qftQECLatAncPrep}{1097376}
\newcommand{\qrcaQECLatDataInteract}{95641}
\newcommand{\qclaQECLatDataInteract}{11921}
\newcommand{\qftQECLatDataInteract}{365792}
\newcommand{\qrcaCritPathPercentData}{5.2}
\newcommand{\qclaCritPathPercentData}{5.3}
\newcommand{\qftCritPathPercentData}{5.0}
\newcommand{\qrcaCritPathPercentInteract}{16.7}
\newcommand{\qclaCritPathPercentInteract}{16.7}
\newcommand{\qftCritPathPercentInteract}{23.7}
\newcommand{\qrcaCritPathPercentPrep}{78.2}
\newcommand{\qclaCritPathPercentPrep}{78.0}
\newcommand{\qftCritPathPercentPrep}{71.2}
\newcommand{\qrcaAvgBWNeedForQECLogical}{34.8}
\newcommand{\qclaAvgBWNeedForQECLogical}{306.1}
\newcommand{\qftAvgBWNeedForQECLogical}{36.8}
\newcommand{\qrcaAvgBWNeedForQECPhysical}{243.5}
\newcommand{\qclaAvgBWNeedForQECPhysical}{2142.5}
\newcommand{\qftAvgBWNeedForQECPhysical}{257.5}
\newcommand{\ancFacSimpleLat}{323}
\newcommand{\ancFacSimpleBWLogical}{3.1}
\newcommand{\qrcaAvgBWNeedForNonTransLogical}{7.0}
\newcommand{\qclaAvgBWNeedForNonTransLogical}{62.7}
\newcommand{\qftAvgBWNeedForNonTransLogical}{8.6}
\newcommand{\qrcaAvgBWNeedForNonTransPhysical}{49.3}
\newcommand{\qclaAvgBWNeedForNonTransPhysical}{438.7}
\newcommand{\qftAvgBWNeedForNonTransPhysical}{60.3}
\newcommand{\stageOneLat}{73}
\newcommand{\stageOneInBWPhysical}{13.7}
\newcommand{\stageOneOutBWPhysical}{13.7}
\newcommand{\stageOneArea}{1}
\newcommand{\stageTwoCNOTLat}{95}
\newcommand{\stageTwoCNOTInBWPhysical}{221.1}
\newcommand{\stageTwoCNOTOutBWPhysical}{221.1}
\newcommand{\stageTwoCNOTArea}{28}
\newcommand{\stageTwoCatLat}{62}
\newcommand{\stageTwoCatInBWPhysical}{96.8}
\newcommand{\stageTwoCatOutBWPhysical}{96.8}
\newcommand{\stageTwoCatArea}{6}
\newcommand{\stageThreeLat}{82}
\newcommand{\stageThreeInBWPhysical}{122.0}
\newcommand{\stageThreeOutBWPhysical}{85.2}
\newcommand{\stageThreeArea}{10}
\newcommand{\stageFourLat}{138}
\newcommand{\stageFourInBWPhysical}{152.2}
\newcommand{\stageFourOutBWPhysical}{50.7}
\newcommand{\stageFourArea}{21}
\newcommand{\stageOneCount}{24}
\newcommand{\stageTwoCNOTCount}{1}
\newcommand{\stageTwoCatCount}{1}
\newcommand{\stageThreeCount}{3}
\newcommand{\stageFourCount}{2}
\newcommand{\stageOneTotalHeight}{24}
\newcommand{\stageTwoCNOTTotalHeight}{4}
\newcommand{\stageTwoCatTotalHeight}{2}
\newcommand{\stageTwoTotalHeight}{6}
\newcommand{\stageThreeTotalHeight}{30}
\newcommand{\stageFourTotalHeight}{42}
\newcommand{\stageOneTotalInBWPhysical}{328.8}
\newcommand{\stageTwoCNOTTotalInBWPhysical}{221.1}
\newcommand{\stageTwoCatTotalInBWPhysical}{96.8}
\newcommand{\stageThreeTotalInBWPhysical}{365.9}
\newcommand{\stageFourTotalInBWPhysical}{304.3}
\newcommand{\stageOneTotalOutBWPhysical}{328.8}
\newcommand{\stageTwoCNOTTotalOutBWPhysical}{221.1}
\newcommand{\stageTwoCatTotalOutBWPhysical}{96.8}
\newcommand{\stageThreeTotalOutBWPhysical}{255.6}
\newcommand{\stageFourTotalOutBWPhysical}{101.4}
\newcommand{\stageOneTotalArea}{24}
\newcommand{\stageTwoCNOTTotalArea}{28}
\newcommand{\stageTwoCatTotalArea}{6}
\newcommand{\stageTwoTotalArea}{34}
\newcommand{\stageThreeTotalArea}{30}
\newcommand{\stageFourTotalArea}{42}
\newcommand{\totalCrossbarArea}{168}
\newcommand{\totalComputeArea}{130}
\newcommand{\totalAncFacArea}{298}
\newcommand{\totalAncFacBWPhysical}{73.5}
\newcommand{\totalAncFacBWLogical}{10.5}
\newcommand{\piEightCatPrepLat}{218}
\newcommand{\piEightCatPrepInBWPhysical}{32.1}
\newcommand{\piEightCatPrepOutBWPhysical}{32.1}
\newcommand{\piEightCatPrepArea}{12}
\newcommand{\piEightTransXSZLat}{53}
\newcommand{\piEightTransXSZInBWPhysical}{264.2}
\newcommand{\piEightTransXSZOutBWPhysical}{264.2}
\newcommand{\piEightTransXSZArea}{7}
\newcommand{\piEightDecodeLat}{218}
\newcommand{\piEightDecodeInBWPhysical}{64.2}
\newcommand{\piEightDecodeOutBWPhysical}{36.7}
\newcommand{\piEightDecodeArea}{19}
\newcommand{\piEightHMTransZLat}{74}
\newcommand{\piEightHMTransZInBWPhysical}{108.1}
\newcommand{\piEightHMTransZOutBWPhysical}{94.6}
\newcommand{\piEightHMTransZArea}{8}
\newcommand{\piEightCatPrepCount}{4}
\newcommand{\piEightTransXSZCount}{1}
\newcommand{\piEightDecodeCount}{4}
\newcommand{\piEightHMTransZCount}{2}
\newcommand{\piEightCatPrepTotalHeight}{24}
\newcommand{\piEightCatPrepTotalInBWPhysical}{128.4}
\newcommand{\piEightCatPrepTotalOutBWPhysical}{128.4}
\newcommand{\piEightCatPrepTotalArea}{48}
\newcommand{\piEightTransXSZTotalHeight}{7}
\newcommand{\piEightTransXSZTotalInBWPhysical}{264.2}
\newcommand{\piEightTransXSZTotalOutBWPhysical}{264.2}
\newcommand{\piEightTransXSZTotalArea}{7}
\newcommand{\piEightDecodeTotalHeight}{52}
\newcommand{\piEightDecodeTotalInBWPhysical}{256.9}
\newcommand{\piEightDecodeTotalOutBWPhysical}{146.8}
\newcommand{\piEightDecodeTotalArea}{76}
\newcommand{\piEightHMTransZTotalHeight}{16}
\newcommand{\piEightHMTransZTotalInBWPhysical}{216.2}
\newcommand{\piEightHMTransZTotalOutBWPhysical}{189.2}
\newcommand{\piEightHMTransZTotalArea}{16}
\newcommand{\piEightTotalCrossbarArea}{256}
\newcommand{\piEightTotalComputeArea}{147}
\newcommand{\piEightTotalArea}{403}
\newcommand{\piEightTotalBWPhysical}{128.4}
\newcommand{\piEightTotalBWLogical}{18.3}
\newcommand{\qrcaDataArea}{679}
\newcommand{\qrcaAvgZeroAncFacArea}{986.9}
\newcommand{\qrcaAvgPiEightAncFacArea}{354.7}
\newcommand{\qrcaAvgTotalAncFacArea}{1341.6}
\newcommand{\qrcaAvgTotalArea}{2020.6}
\newcommand{\qrcaAvgAreaPercentData}{33.6}
\newcommand{\qrcaAvgAreaPercentZeroAncFac}{48.8}
\newcommand{\qrcaAvgAreaPercentPiEightAncFac}{17.6}
\newcommand{\qclaDataArea}{861}
\newcommand{\qclaAvgZeroAncFacArea}{8682.2}
\newcommand{\qclaAvgPiEightAncFacArea}{3154.4}
\newcommand{\qclaAvgTotalAncFacArea}{11836.7}
\newcommand{\qclaAvgTotalArea}{12697.7}
\newcommand{\qclaAvgAreaPercentData}{6.8}
\newcommand{\qclaAvgAreaPercentZeroAncFac}{68.4}
\newcommand{\qclaAvgAreaPercentPiEightAncFac}{24.8}
\newcommand{\qftDataArea}{224}
\newcommand{\qftAvgZeroAncFacArea}{1043.5}
\newcommand{\qftAvgPiEightAncFacArea}{433.7}
\newcommand{\qftAvgTotalAncFacArea}{1477.1}
\newcommand{\qftAvgTotalArea}{1701.1}
\newcommand{\qftAvgAreaPercentData}{13.2}
\newcommand{\qftAvgAreaPercentZeroAncFac}{61.3}
\newcommand{\qftAvgAreaPercentPiEightAncFac}{25.5}

\newcommand{\sanctified}{High-Fidelity }  % this word is used for encoded ancilla coming out of the factory

\thispagestyle{empty}

\begin{abstract}

{\it We analyze circuits for a number of kernels from popular quantum
computing applications, characterizing the hardware resources
necessary to take ancilla preparation off the critical path.  The
result is a chip entirely dominated by ancilla generation circuits.
To address this issue, we introduce optimized ancilla factories and
analyze their structure and physical layout for ion trap technology.
We introduce a new quantum computing architecture with highly
concentrated data-only regions surrounded by shared ancilla factories.
The results are a reduced dependence on costly teleportation, more
efficient distribution of generated ancillae and more than five times
speedup over previous proposals.}

\end{abstract}

\section{Introduction}\label{sec:intro}

Quantum computing shows great potential to speed up difficult
 applications such as factorization~\cite{shor1995srd} and quantum
 mechanical simulation~\cite{zalka1998sqs}. Unfortunately, quantum
 states are so fragile that all quantum bits, or {\it qubits}, in the
 system must be encoded for redundancy and remain encoded during
 computation.  Various encoding methodologies have been
 proposed~\cite{calderbank1996gqe, steane2003oan}, ranging from several
 to several dozen {\it physical qubits} used to represent a single {\it
 encoded qubit} to be used in the high-level computation.

It is expected that an encoded qubit will need to undergo a Quantum
Error Correction (QEC) step after each ``useful'' basic gate is
performed upon it.  However, the bulk of a QEC operation is a
preparation circuit involving the creation of encoded ancillary
qubits, or {\it ancillae}, which does not involve the data qubit to be
corrected.  Consequently, as Chi et al. point out
in~\cite{chi2007tqa}, the critical path of a quantum circuit could be
significantly reduced if the ancilla preparation work were done in
parallel with useful computation.  In particular, the speed of a
quantum computation would be limited solely by \emph{data
dependencies} between encoded qubits.  We refer to this fully offline
parallelization of data-independent work as {\it running the circuit
at the speed of data}.

\begin{figure}
\begin{center}
\epsfig{file=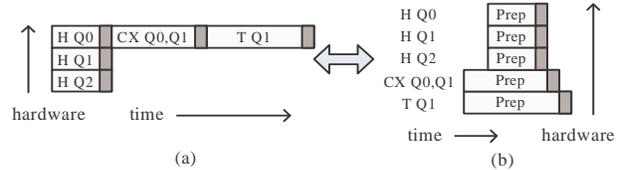,width=\hsize}
\end{center}
\vspace{-0.3in}
\caption{(a) Standard implementation of a circuit involving qubits Q0,
  Q1 and Q2. Only the grey blocks represent interactions with actual
  data. The bulk of the critical path involves independent ancilla
  preparation. (b) An optimized version of the circuit in which ancilla
  preparation is pulled off the critical path through use of increased
  hardware.  Here, the speed of the computation is limited only by data
  dependencies (grey blocks).  }
\label{fig:crit_path}
\end{figure}

Figure~\ref{fig:crit_path}a shows a possible execution of a simple
series of quantum gates involving qubits Q0, Q1 and Q2.  Each gate
involves some encoded ancilla preparation for the QEC step which must
follow it.  In addition, some gates, called {\it non-transversal
gates}, require further encoded ancilla preparation simply to be
performed (elaborated upon in Section~\ref{sec:pi8circuit}).
Figure~\ref{fig:crit_path}b shows these operations performed at the
speed of data.  Chi et al. suggest that these ancilla preparation
operations could be done in advance, but the hardware cost for this
parallelization grows quickly as the critical path is shortened.

In Section~\ref{sec:prepare}, we investigate quantum circuits for
encoded ancilla preparation and evaluate them in terms of error and
complexity.  In Section~\ref{sec:motivation}, we identify three common
subcircuits of larger quantum algorithms and evaluate their
characteristics concerning encoded ancilla needs for both QEC and
non-transversal gates.  In Section~\ref{sec:factory}, we detail the
layout and throughput of a pipelined {\it ancilla factory} specialized
for generating encoded ancilla qubits.  In Section~\ref{sec:arch}, we
combine our analyses to answer the overall question of the feasibility
of running a quantum circuit at the speed of data, and we conclude in
Section~\ref{sec:conclusion}.

%% Intro should include figure showing quantum critical paths and effect
%% of taking encoded ancilla prep off crit path: should also obviously
%% explain what encoded ancillae are; include mention of zero vs non-zero

%% logical (or encoded?) vs. physical: necessary for encoding

%% gates available: in 2.1

%% QEC = quantum error correction

%% ``running at the speed of data''

%% Our contributions:
%% \begin{enumerate}
%% \item analyses of real circuit needs for running at speed of data
%% \item pipelined ancilla factory: great way to keep ancilla
%% distribution ballistic by focusing output on data region
%% \item microarch/floorplan for ion traps: Figure~\ref{fig:arch_final}
%% \begin{itemize}
%% \item compute region contains {\it only} data: allowed ballistic
%% communication: less teleportation -> smaller teleport interconnect
%% \item entire compute region served by multiple ancilly factories:
%% encoded ancilla distributed as needed, not wasted on idle data
%% \item ancillae moved ballistically
%% \item data teleported between compute regions
%% \end{itemize}
%% \end{enumerate}

\section{Ancilla Preparation Circuits}\label{sec:prepare}

%% {\bf
%% \begin{itemize}
%% \item Figure~\ref{fig:adder_microarch}: 3 blocks: ancilla prep, data,
%% interconnect
%% \item argument: relative area depends on ancilla gen throughput: let's
%% estimate what HW resources it would take to run at speed of data (cite
%% Eric Chi)
%% \item data needs and ancilla generation handled in this section;
%% interconnect in next section (putting it all together)
%% \end{itemize}
%% }

Typical quantum circuits require many encoded ancilla qubits. In this
section, we discuss several ancilla preparation circuits and evaluate
them in terms of complexity and error. Ultimately, we select encoding
circuits that will be used in our layouts in Section~\ref{sec:factory}.

\subsection{Computing on Encoded Data Bits}\label{sec:qec_intro}

\begin{figure}
\begin{center}
\epsfig{file=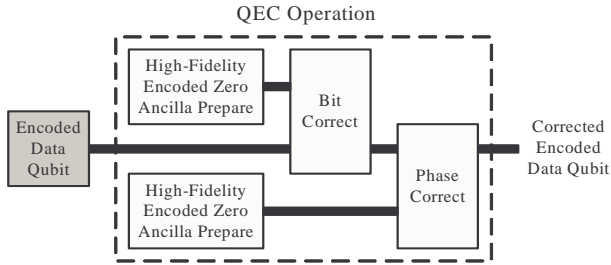,width=\hsize}
\end{center}
\vspace{-0.2in}
\caption{A quantum error correcting (QEC) operation is composed of a
{\it bit-flip} correction and a {\it phase-flip} correction,
corresponding to the two types of errors that can happen to a qubit.
The thick bars represent encoded qubits.}
%``Z correction'' and an ``X correction'' corresponding to two
%orthogonal vectors of errors that can happen to a qubit.  A Z error is
%also known as a \emph{phase flip} and a X error is also known as a
%\emph{bit flip}.}
\label{fig:basic_ecc}
\end{figure}

Since quantum data is very fragile, it must be encoded at all times in
an appropriate quantum error correction code.  
%\subsection{Quantum Error Correction}\label{sec:qec_intro}
%\subsection{Fault Tolerant Circuits}
A high-level view of the procedure for error-correcting an encoded data
qubit is shown in Figure~\ref{fig:basic_ecc}.  Both the \emph{bit value}
and \emph{phase} must be repaired during the QEC
step~\cite{nielsen2000qca}. Two sets of physical ancilla qubits are each
encoded into the zero state and then consumed during correction.

%% We can break down an error correction procedure into three steps as
%% shown in Figure \ref{fig:basic_ecc}.  The reason we have to extract
%% parity information into the ancillae before measurement instead of
%% directly measuring the data, is due to the fact that the data is in a
%% superposition of many possible values and direct measurement would
%% collapse the state and invalidate the computation.  We must do an
%% ``X'' and ``Z'' correction because in quantum systems, we must correct
%% both bit and phase flip errors \cite{???}.

\begin{figure*}
\begin{center}
\epsfig{file=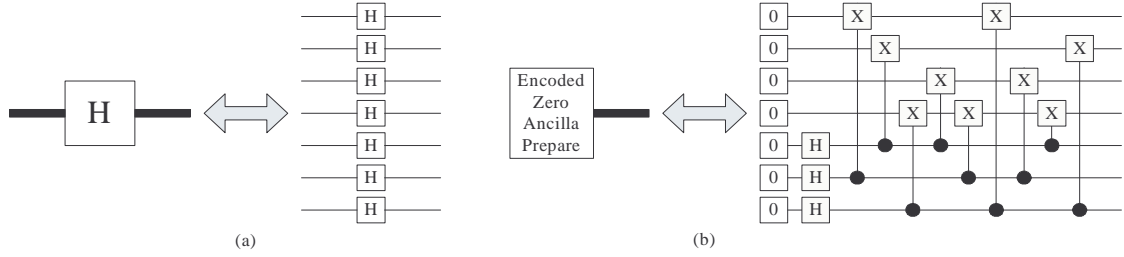,width=0.9\hsize}
\end{center}
\vspace{-0.2in}
\caption{(a) A {\it transversal} encoded gate involves transversal
application of physical gates.  (b) A {\it non-transversal} encoded
gate involves multi-qubit physical operations between physical qubits
within the same encoded qubit.}
% decompose into physical gates that only
%  operate on single qubits like the Hadamard gate on the left.  The
%  zero preparation module on the right is non-transversal since there
%  are CNOT gates between qubits within the block.}
\label{fig:trans_vs_nontrans}
\end{figure*}

Gates applied to encoded data may be classified into two types: {\it
transversal} and {\it non-transversal}.  A transversal encoded gate is
applied by performing the corresponding physical gate
\emph{independently} on each of the qubits comprising the encoded
qubit, as shown in Figure~\ref{fig:trans_vs_nontrans}a for the
Hadamard gate.  A non-transversal encoded gate is decomposed into a
more complex set of physical operations, including multi-qubit
physical operations between physical qubits within the same encoded
qubit; for example, see the Basic Encoded Zero Ancilla Prepare in
Figure~\ref{fig:trans_vs_nontrans}b.  Since errors are propagated
between physical qubits during the application of non-transversal
gates, such gates must be designed carefully to avoid introducing
uncorrectable errors.

%% Because error rates are so high in quantum circuits, we must be very
%% careful to limit the propagation of errors through the circuit.  Most
%% of the paranoia associated with fault tolerance in quantum circuits
%% \cite{preskill1997ftq} involves limiting the propagation of errors
%% through 2 qubit gates like CNOTs.  The central rule for fault
%% tolerance is that our circuits may never allow more gate mediated
%% qubits correlations inside an encoded block than the number of errors
%% the code can correct as shown in Figure
%% \ref{fig:transversal_vs_nontransversal}.  Otherwise, qubit errors will
%% propagate too fast for any amount of error correction to correct.

A class of quantum codes known as CSS codes \cite{steane1996mpi,
calderbank1996gqe} allow transversal implementations of most encoded
gates.  For this reason, CSS codes are used in most analyses of the
fault tolerance of quantum circuits.  For the rest of this paper, we
use the [[7,1,3]] CSS code \cite{steane1996mpi}.
%which can only correct a single bit, so having transversal gates is
%critical for designing a fault tolerant circuit using this code.
Encoded gates that can be performed transversally on this code include the
two-qubit CX, as well as the one-qubit X, Y, Z, Phase, and Hadamard
gates.  In order to have a universal gate set, we also need the
non-transversal $\pi/8$ gate and the encoding procedure to create an
encoded ancilla.  We will discuss how to obtain a fault tolerant
version of the $\pi/8$ gate later in this section, but first we
investigate the problem of getting a fault tolerant encoding
procedure.

\subsection{Circuit Evaluation Methodology} \label{sec:anc_method}

Since encoded ancillae are a major component of error correction, it
is critical to generate \emph{clean} ancillae to avoid introducing
errors during the correcting process. In the following, we will
evaluate circuits by using the tools in \cite{whitney2007agl} which
allow us to lay out circuits. The effects of error are then modeled by
Monte Carlo simulation where errors can be introduced at any gate or
qubit movement operation.  Additionally, we model the fact that
two-qubit gates propagate bit and phase flips between qubits.  This
simulation is similar to what was done in \cite{steane2003oan} except
with the addition of qubit movement error from our detailed layout.
We assume an independent error probability for each gate and movement
operation.  The gate error rate is $10^{-4}$ and the error per
movement op is $10^{-6}$.  Our gate and movement error rates are
consistent with \cite{steane2004bbg}.

\subsection{Encoded Ancilla Preparation}\label{sec:ftsp}

Since the Bit Correct and Phase Correct circuits in Figure
\ref{fig:basic_ecc} are fully transversal (each consisting of a
transversal CX, measure and conditional
correct~\cite{preskill1997ftq}), we focus on the basic zero ancilla
preparation circuit, shown in Figure~\ref{fig:trans_vs_nontrans}b.
The probability of an uncorrectable error in the resulting encoded
output of this circuit is $1.8 \times 10^{-3}$ based on our evaluation
methodology above.  We would like to improve on this basic result.

There are two different circuit-level techniques for removing general
errors from an encoded qubit: {\it verification} and {\it correction}.
Verification tests a qubit in a known state for error and discards it
if too much error is found.  Correction is more complex, but it
corrects a bit or phase error from an encoded qubit in an unknown
state, thus it is more suitable for data qubits in a long-running
computation.  Encoded zero ancillae are in known state and may be
discarded if necessary, so either method is suitable.

\begin{figure*}
\begin{center}
\epsfig{file=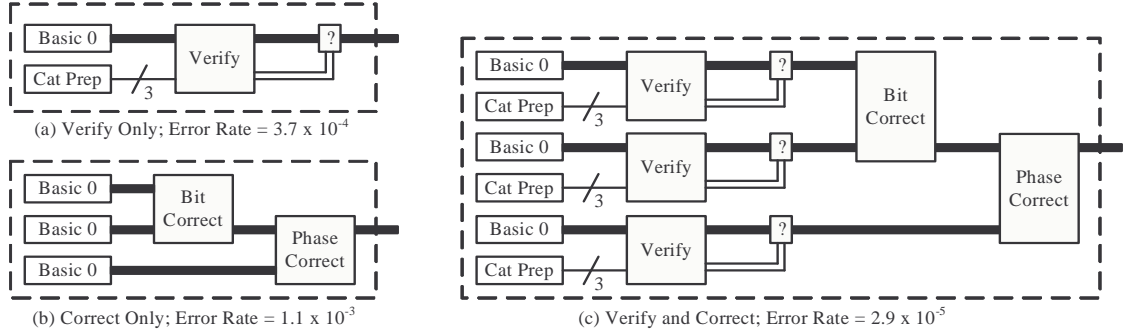,width=0.9\hsize}
\end{center}
\vspace{-0.2in}
\caption{Different circuits for the ``\sanctified Encoded Zero
Ancilla Prepare'' in Figure~\ref{fig:basic_ecc}.  Each ``Basic 0''
module corresponds to the circuit in
Figure~\ref{fig:trans_vs_nontrans}b.  Each ``Cat Prep'' module
corresponds to the preparation of a special 3-qubit state.  Thick bars
are encoded qubits (seven physical qubits).  The overall error rate of
each is given under each circuit.}
%% Different implementations of the ``Clean 0 Prep'' module from
%%   Figure \ref{fig:basic_ecc}.  The ``0 Prep'' block is the same as the
%%   circuit in Figure \ref{fig:transversal_vs_nontransveral}b.  Circuit
%%   a) is the simplest design, here we do not try to filter error/s at
%%   all, the probability of there being an error in the block is $1.8
%%   \times 10^{-3}$.  Circuit b) uses a verification stage to filter
%%   out exceptionally erroneous states and has an error rate of $3.7
%%   \times 10^{-4}$.  Circuit c) uses a correction stage to fix any
%%   problems based on parity checks to other identical ancilla and has
%%   an error rate of $1.1 \times 10^{-3}$.  Circuit d) first filters out
%%   errors with verification on all ancilla blocks which then are used
%%   to correct out additional errors.  The error rate of this is $2.9
%%   \times 10^{-5}$.}
\label{fig:anc_prep_options}
\end{figure*}

While Figure~\ref{fig:trans_vs_nontrans}b shows the circuit for
preparing an encoded ancilla in the zero state in the [[7,1,3]] CSS
code, we would like a more error-free ancilla qubit for interaction
with data.  Figure~\ref{fig:anc_prep_options} shows some example zero
ancilla preparation circuits from the literature \cite{svore2006ntf,
preskill1997ftq}, with the overall error rate for each given under
the circuit.  Correction alone (Figure~\ref{fig:anc_prep_options}b)
loses to verification alone (Figure~\ref{fig:anc_prep_options}a) in
both error and area.  When comparing
Figures~\ref{fig:anc_prep_options}a and ~\ref{fig:anc_prep_options}c,
it is important to note that they are not to scale.  The ``Basic 0''
module (expanded in Figure~\ref{fig:trans_vs_nontrans}b) is by far the
most complex, so by doing both verification and correction, we get
more than an order of magnitude improvement in error over verification
alone for slightly more than three times the area.  Thus, we shall use
the circuit in Figure~\ref{fig:anc_prep_options}c in this paper.

Since we are using qubit verification as part of our encoded zero
preparation, we need to know the success rate of verification.  Using
the same Monte Carlo simulation used for error probability
calculations, we estimate the verification failure rate of the subunit
\ref{fig:anc_prep_options}a to be 0.2\%.  We will use this in
calculations later in Section \ref{sec:pipe_fac}.

\subsection{Fault Tolerant $\pi/8$ Gate}\label{sec:pi8circuit}

\begin{figure*}
\begin{center}
\epsfig{file=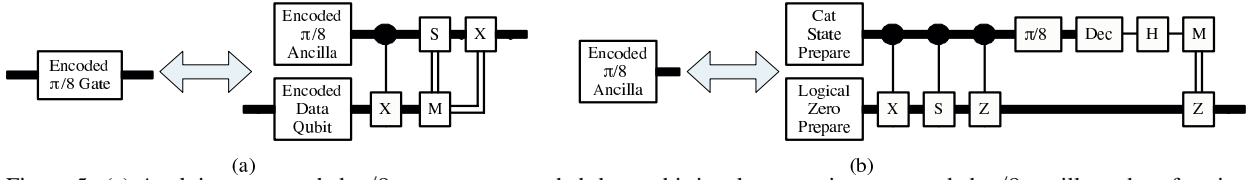,width=\hsize}
\end{center}
\vspace{-0.3in}
\caption{(a) Applying an encoded $\pi/8$ gate on an encoded data qubit
involves creating an encoded $\pi/8$ ancilla and performing some
transversal gates.  (b) Creating the encoded $\pi/8$ ancilla used in
the circuit in (a) requires an encoded zero ancilla, a 7-qubit cat
state (a specially prepared qubit set) and a series of transversal
gates.  Note that the $\pi/8$ gate near the far right is transversal
but does not implement an encoded $\pi/8$ gate.}
%% Since a $\pi/8$ gate has no transversal implementation on a
%%   [[7,1,3]] encoded block, a fault tolerant implementation is more
%%   complex.  For fault tolerance, we require the creation of a special
%%   $\pi/8$ state followed by other transversal gates interacting with
%%   the data (left).  Creating the special state (right), requires
%%   clean zero and cat ancilla states followed by transversal gates
%%   between them.  Note that we are not cheating, the $\pi/8$ gate on
%%   the far right is just a straightforward single qubit transversal
%%   gate, but it does not perform an effective $\pi/8$ rotation on the
%%   encoded state.}
\label{fig:pi8_anc_prep}
\vspace{-0.1in}
\end{figure*}

It has been shown that {\it no} quantum error correcting code has
transversal gate implementations for all the gates in a universal set
\cite{zeng2007tvu}, and indeed, in the [[7,1,3]] CSS code, we need the
non-transversal $\pi/8$ gate in order to complete the universal set.
In order to maintain fault tolerance when performing the $\pi/8$ gate
on a [[7,1,3]] encoded qubit, we use a technique developed in
\cite{zhou2000mql}.  Their approach is to generate an encoded ancilla
qubit encoded in the $\pi/8$ state and perform transversal
interactions with the data, as shown in
Figure~\ref{fig:pi8_anc_prep}a, to achieve the overall effect of an
encoded $\pi/8$ gate.

To encode the $\pi/8$ ancilla qubit, we could try to create a physical
$\pi/8$ ancilla qubit and then use the encoding circuit in
Figure~\ref{fig:trans_vs_nontrans}b, but this would result in errors
on the original physical qubit propagating to each physical qubit in
the final encoded ancilla, which is unacceptable.  Thus, we require
the far more complicated circuit shown Figure~\ref{fig:pi8_anc_prep}b,
which consists of an encoded zero ancilla prepare, a 7-qubit cat state
prepare (where a cat state is a specially prepared multi-qubit state)
and a series of transversal encoded gates.

%%   Their method involves the creation of a special
%% ancilla state which the data is teleported into, followed by
%% additional transversal gates to adjust the resulting state (Figure
%% ???).  Since the teleportation component of the circuit involves only
%% transversal CNOTs and measurements, the only tricky part that remains
%% is the creation of the special ancilla state.

%% Since we know what the special ancilla state is, we would be tempted
%% to performs a circuit as shown in Figure \ref{fig:pi8_anc_prep}a and
%% then perform correction or verification operations to remove all
%% possible errors.  This does not work though, since the qubit
%% undergoing the rotation is unencoded, the later encoding does not
%% recognize any potential gate errors from the unencoded gate.  This
%% leads to arbitrary errors in the resulting encoded $\pi/8$ ancilla
%% state.  Instead, the process follows Figure \ref{fig:pi8_anc_prep}b.
%% Creation of the special ancilla $\pi/8$ state starts with an encoded
%% zero state which is presumably produced using the methods outlined in
%% Section \ref{sec:ftsp}.  It then performs transversal operations with
%% a cat ancilla state, the cat state is measured which projects the
%% encoded qubits to the proper $\pi/8$ ancilla state.

%% We wrap up the entire circuit in Figure \ref{fig:pi8_anc_prep}b into
%% a \emph{$\pi/8$ ancilla factory} which is composed of a zero ancilla
%% factory and a cat state factory.

\subsection{Fault Tolerant $\pi/2^k$ Gates}\label{sec:pi2k}

The Quantum Fourier Transform (QFT) requires controlled phase rotation
gates by small angles (these gates replace the explicit tracking of
roots of unity in the classical FFT algorithm).  The amount of
precision for these gates scales exponentially in the number of bits
involved in the QFT \cite{nielsen2000qca}.  A controlled phase
rotation by $\pi/2^k$ can be generated by a CX gate and 3 single
qubit $\pi/2^{k+1}$ gates \cite{fowler2005tls}.  Thus, using circuit
techniques mentioned so far, we can implement every gate in the QFT
fault tolerantly except these single qubit rotation gates.  There are
two problems with implementing an arbitrary precision phase rotation
fault tolerantly:
\begin{itemize}
\item For angles smaller than $\pi/2$, there is no transversal gate
  implementation using the [[7,1,3]] code \cite{zeng2007tvu}.  In
  fact, this seems likely to be true for all codes.
\item Such a gate would require the physical implementation of an
  arbitrary precision rotation --  a difficult burden on the
  engineers of these devices.
\end{itemize}
%% \cite{barenco1995egq} showed that it is possible to approximate an
%% arbitrary phase rotation using Hadamard and $\pi/8$ gates.  The
%% Solovay-Kitaev theorem \cite{kitaev1997qca} tells us that this
%% generation gate sequence is polylogarithmic in the amount of error we
%% can tolerant in the precision.  Since we already discussed how to
%% implement fault tolerant Hadamard and $\pi/8$ gates, we know how to
%% make these gate sequences fault tolerant.  However, these proofs are
%% not constructive so t
Due to the above reasons, we adopt a technique by Fowler
\cite{fowler2005tls}.  To approximate small angle rotations, we
exhaustively search all permutations of T and H gates to find a
minimum length sequence for a $\pi/2^k$ rotation gate up to an
acceptable error.

\begin{figure}
\begin{center}
\epsfig{file=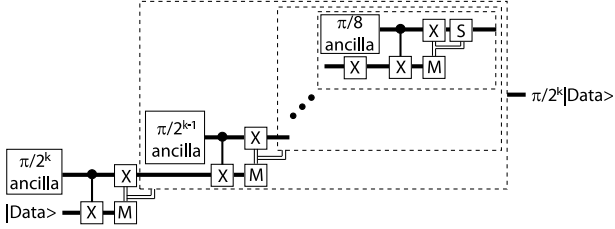,width=\hsize}
\end{center}
\vspace{-0.2in}
\caption{Fault tolerant $\pi/2^k$ gates can be performed recursively
  with a cascade of $\pi/2^i | i = {3...k}$ ancilla factories and
  $k-2$ CX and X gates.  Each measure gate output controls both the
  single qubit X gate and the compound gate involving more ancilla
  factories.  Each measurement has a equal chance of giving the
  ``correct'' state, in which the remaining circuit is skipped or a
  ``wrong'' state in which a larger rotation has to be done to adjust
  the state.  The actual output data from the circuit connects to the
  first quantum bitline associated with a correct measurement.}
\label{fig:pi_n_anc_prep}
\end{figure}

We also note that if a $\pi/2^k$ physical gate is available in a given
technology, an exact fault-tolerant $\pi/2^k$ can be implemented as
shown in Figure \ref{fig:pi_n_anc_prep}.  In order to be conservative
about the availability of arbitrary precision rotation gates, we do
not use this construction in the circuits in this paper.  However, in
Section \ref{sec:pi8factory}, we briefly analyze the performance
advantages of this technique.

%Additionally, work done in \cite{fowler2005tls} shows that the
%accuracy of Shor's algorithm, using the QFT, is not significantly
%degraded by eliminating all rotation gates smaller than $\pi/128$ for
%transforms of up to 1024 bits.  Our QFT benchmark operates on 32 bits
%and therefore we conservatively use this approximate QFT approach of
%only using rotation gates up to $\pi/128$.  This approach results in a
%valid QFT circuit, but one which exhibits less parallelism than one
%might expect from the Fourier Transform.  We are currently
%investigating alternate approaches to a QFT circuit.

\section{Circuit Characteristics}\label{sec:motivation}

We now characterize the runtime properties of some commonly used
quantum circuits, focusing on the impact of encoded ancilla
generation.  Many quantum algorithms require ancillae to assist in
computation.  For example, an $n$-bit Quantum Ripple-Carry Adder uses
two $n$-bit data inputs plus $n+1$ ancillae.  In addition to this,
shorter-lived ancillae are needed for QEC and for performing
non-transversal encoded gates, as discussed earlier.

%For the remainder of 
Throughout this paper we refer to the longer-lived ancillae used in
the main computation as ``data ancillae'' and to the shorter-lived
ones as ``ancillae.''  We make this distinction because data ancillae
tend to have long enough lifespans that ``discarding'' them and
restarting their portion of the computation has a relatively high
cost.  Our work focuses on the short-lived ancillae which need to be
produced in large quantities and which may more easily be discarded
and re-encoded.
% if they are found to be faulty.

%\begin{table}
%\small
%\begin{minipage}{0.45\hsize}

\begin{table}
\begin{center}
\begin{tabular}{|l|c|c|}
\hline
Physical & Latency & Latency \\
Operation & Symbol & ($\mu$s) \\
\hline \hline
One-Qubit Gate & $t_{1q}$ & \tOneQGate \\
Two-Qubit Gate & $t_{2q}$ & \tTwoQGate \\
Measurement & $t_{meas}$ & \tMeas \\
Zero Prepare & $t_{prep}$ & \tPhysPrep \\
\hline
\end{tabular}
\end{center}
\vspace{-0.2in}
\caption{The latency values for various physical operations in ion
trap technology~\cite{steane2004bbg,thaker2006qmh,isailovic2006ins}.
Since these change as more experiments are done, we show many of our
results in a symbolic fashion before plugging in these values.}
\label{table:ion_trap_gate_stats}
\end{table}
%\end{minipage}
%\hfill
%\begin{minipage}{0.45\hsize}
% The Qubit Count table used to be here in a minipage.
%\end{minipage}
%\end{table}
We do most of our
%We will try to do much of our 
analysis in a symbolic fashion so that
it may be applied to varying technologies and assumptions.  However,
we will also be applying the analysis to a specific 
%example
technology, trapped ions~\cite{seidelin06}, in order to make the
results of our calculations more concrete.  We use the
%We will be using the
physical gate latencies shown in
Table~\ref{table:ion_trap_gate_stats}, the [[7,1,3]] CSS code
introduced in Section~\ref{sec:qec_intro} and the encoded ancilla
preparation circuits shown in Figures~\ref{fig:anc_prep_options}c and
\ref{fig:pi8_anc_prep}b.  Note that the ``Zero Prepare'' in
Table~\ref{table:ion_trap_gate_stats} refers to a physical zero
prepare, which is the leftmost set of gates in the Basic Encoded Zero
Ancilla Prepare in Figure~\ref{fig:trans_vs_nontrans}b.

%% In the next section, we discuss the three benchmarks we have chosen to
%% study.  Then in the following section, we characterize various properties
%% of these quantum circuits, which shall aid us in making design decisions
%% in the remainder of the paper.

\subsection{Benchmarks} \label{sec:benchmarks}

% This table technically should go in the benckmarks subsection.
%% \begin{table}
%% \begin{center}
%% \begin{tabular}{|l|c|c|}
%% \hline
%% Circuit & Qubit Count & Gate Count \\
%% \hline \hline
%% 32-bit QRCA \cite{draper2000aqc} & \qrcaNumQubits & \qrcaTotalGateCount \\
%% 32-bit QCLA \cite{draper2004ldq} & \qclaNumQubits & \qclaTotalGateCount \\
%% 32-bit QFT \cite{nielsen2000qca} & \qftNumQubits & \qftTotalGateCount \\
%% \hline
%% \end{tabular}
%% \end{center}
%% \vspace{-0.2in}
%% \caption{Total encoded qubit count and total encoded gate count for
%% our three benchmarks, the 32-bit Quantum Ripple-Carry Adder (QRCA),
%% the 32-bit Quantum Carry-Lookahead Adder (QCLA) and the 32-bit Quantum
%% Fourier Transform (QFT).}
%% \label{table:circuit_sizes}
%% \end{table}

For our benchmarks, we use the 32-bit Quantum Ripple-Carry Adder
(QRCA) circuit from \cite{draper2000aqc}, the 32-bit Quantum
Carry-Lookahead Adder (QCLA) circuit from \cite{draper2004ldq} and a
32-bit Quantum Fourier Transform (QFT) circuit we derived using
methodology described in Section~\ref{sec:pi2k}.  All three
%Both the QFT and the adders
are core kernels of a varied array of quantum algorithms, including
Shor's factorization algorithm.

\subsection{QEC Circuit Characteristics}\label{sec:circuit_needs}

%% {\bf
%% \begin{itemize}
%% \item FIG (Figure~\ref{fig:adding_qec}) and EQUATIONS
%% (\ref{eq:latnoqec} and~\ref{eq:latwithqec}): time devoted to ancilla
%% prep for sample circuit
%% \item cite Eric Chi's paper
%% \item TABLE (Table~\ref{table:qeclatenciesbyadder}): time devoted to
%% ancilla prep for each adder
%% \item TABLE (Table~\ref{table:qeclatenciesbyencoding}): time devoted
%% to ancilla prep for multiple encodings
%% \item TABLE (Table~\ref{table:anccountbyadder}): ancilla count by type
%% for each adder
%% \item GRAPH (Figures~\ref{fig:ancneedsripple}, \ref{fig:ancneedscla}
%% and \ref{fig:ancneedscsl}): ancilla needs over time for each adder
%% \item GRAPH (Figure~\ref{fig:ancavailripple}): ancilla BW available vs
%% execution time
%% \end{itemize}
%% }

%% As mentioned earlier, Chi et al. \cite{???} point out that the data
%% qubit being corrected is only directly involved in a minor portion of
%% each QEC step, meaning that the rest of the QEC step (largely the
%% encoded ancilla preparation) could be taken off the critical path
%% (also shown in Table~\ref{table:qec_latencies}).  Of course, supplying
%% encoded ancillae at this increased rate would require far more
%% hardware resources, as discussed further in Setion~\ref{???}.  But
%% first,

We study our benchmark circuits at two extremes of the latency-area
trade-off: 1) No overlap of QEC and computation
% as in Figure~\ref{fig:adding_qec}b
(high latency, but low area), and 2) infinitely fast encoded ancilla
production, resulting in an execution limited only by data
dependencies (low latency, but potentially much higher area for
encoded ancilla generation).

\begin{table*}
\small
\begin{center}
  \begin{tabular}{|l|c|c|c|}
    \hline
     & \multicolumn{1}{|c|}{Data Op Latency ($\mu$s)}
     & \multicolumn{1}{|c|}{Data QEC Interact Latency ($\mu$s)}
     & \multicolumn{1}{|c|}{Ancilla Prep Latency ($\mu$s)} \\
    Circuit
     & \multicolumn{1}{|c|}{(\% of total)}
     & \multicolumn{1}{|c|}{(\% of total)}
     & \multicolumn{1}{|c|}{(\% of total)} \\
    \hline \hline
    32-Bit QRCA & \qrcaCritPath\ (\qrcaCritPathPercentData\%)
     & \qrcaQECLatDataInteract\ (\qrcaCritPathPercentInteract\%)
     & \qrcaQECLatAncPrep\ (\qrcaCritPathPercentPrep\%) \\
    32-Bit QCLA & \qclaCritPath\ (\qclaCritPathPercentData\%)
     & \qclaQECLatDataInteract\ (\qclaCritPathPercentInteract\%)
     & \qclaQECLatAncPrep\ (\qclaCritPathPercentPrep\%) \\
    32-Bit QFT & \qftCritPath\ (\qftCritPathPercentData\%)
     & \qftQECLatDataInteract\ (\qftCritPathPercentInteract\%)
     & \qftQECLatAncPrep\ (\qftCritPathPercentPrep\%) \\
    \hline
  \end{tabular}
\end{center}
\vspace{-0.2in}
\caption{Relative latency of useful data operations, interaction of
data with encoded ancillae for QEC and encoded ancilla preparation for
QEC for various circuits, assuming no overlap between them.}
\label{table:qeclatenciesbyadder}
\end{table*}

\begin{figure*}
\begin{minipage}{0.33\hsize}
\begin{center}
\epsfig{file=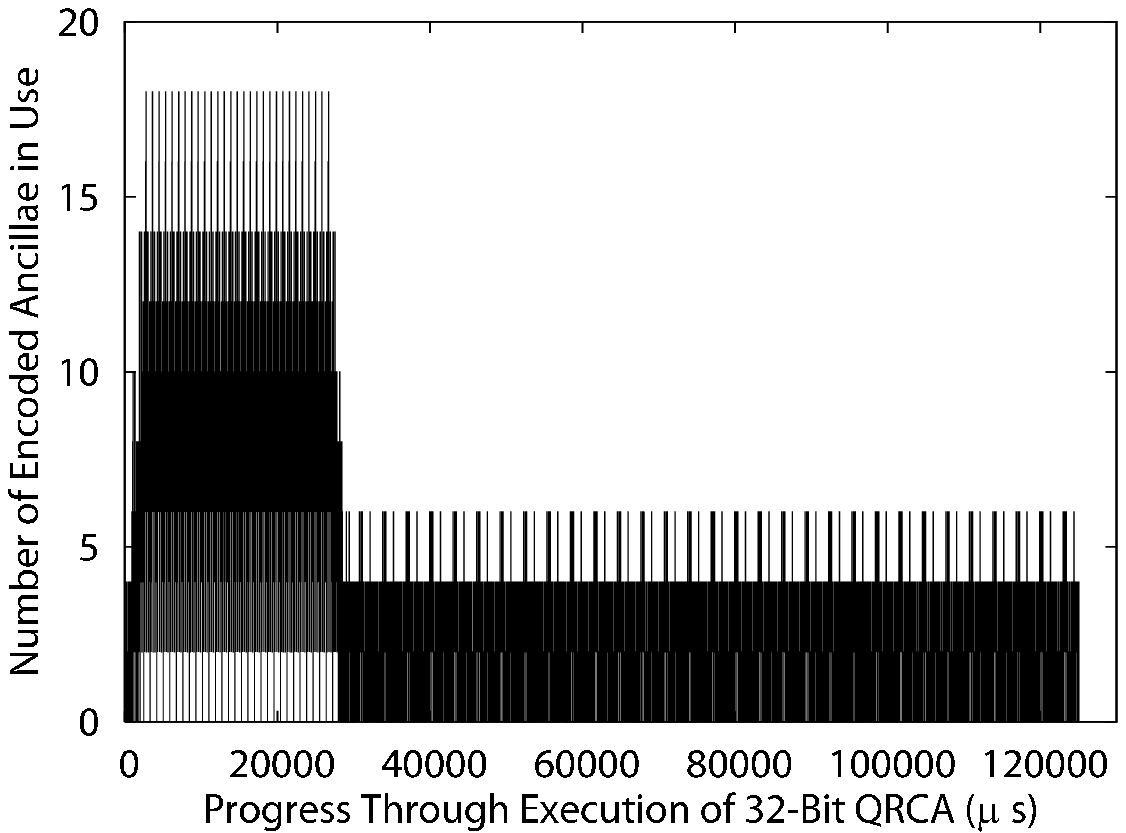,width=\hsize}
\end{center}
%\vspace{-0.2in}
%\caption{Encoded zero ancilla needs for the QRCA if it is run at the
%speed of data.}
\label{fig:ancNeedsQRCA}
\end{minipage}%
\hfill%
\begin{minipage}{0.34\hsize}
\begin{center}
\epsfig{file=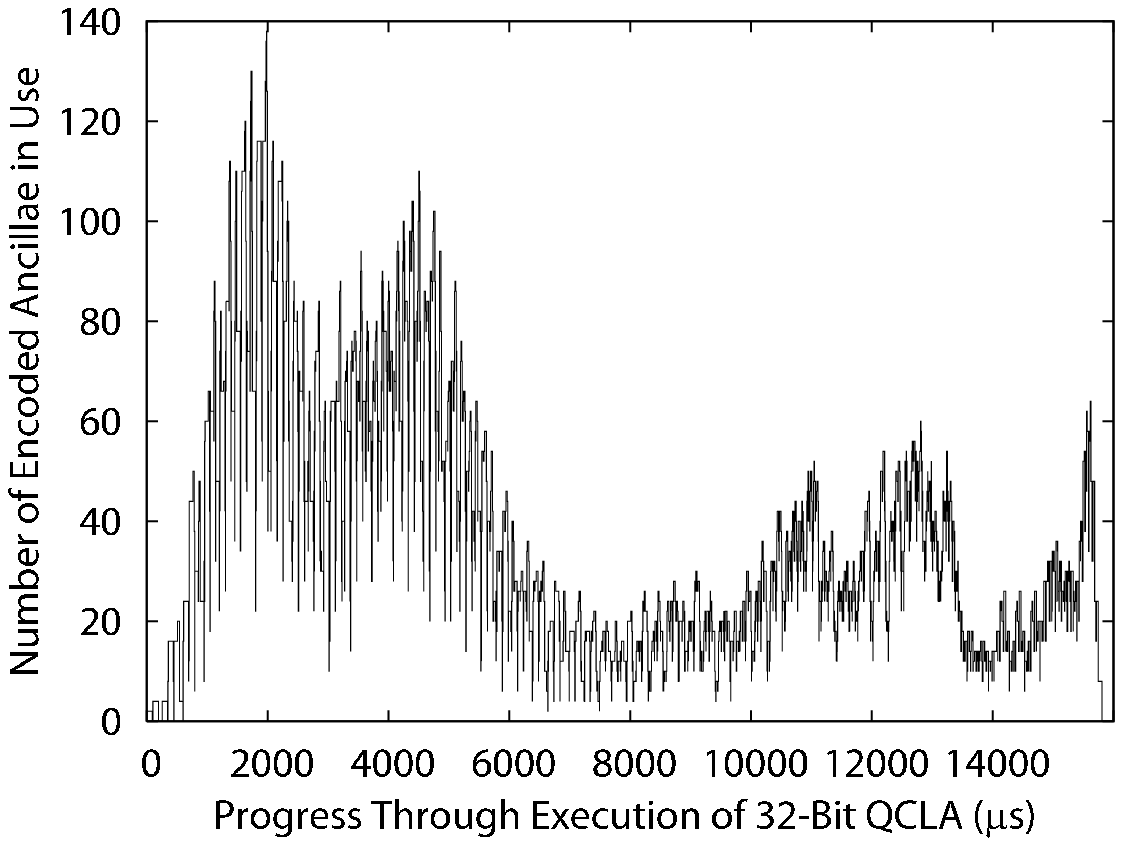,width=\hsize}
\end{center}
%\vspace{-0.2in}
%\caption{Encoded zero ancilla needs for the QCLA if it is run at the
%speed of data.}
\label{fig:ancNeedsQCLA}
\end{minipage}%
\hfill%
\hspace{-0.1in}
\begin{minipage}{0.33\hsize}
\begin{center}
\epsfig{file=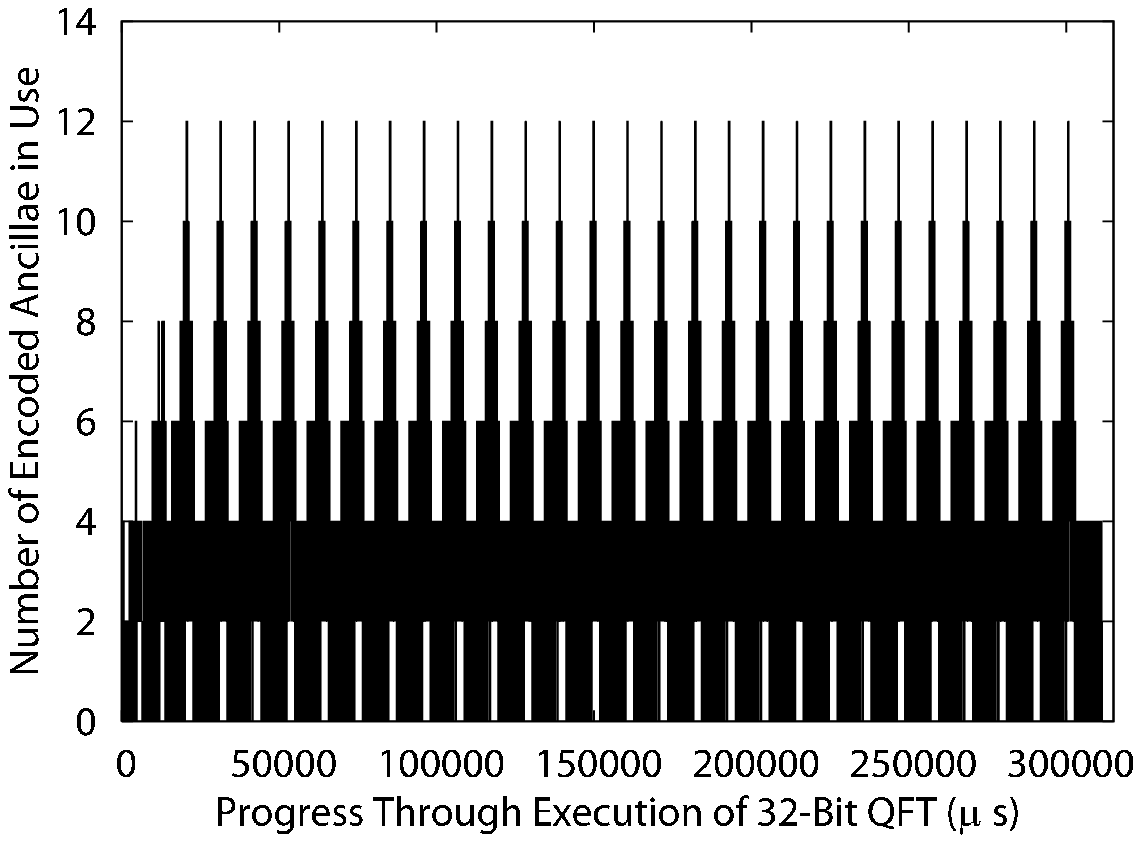,width=\hsize}
\end{center}
%\vspace{-0.2in}
%\caption{Encoded zero ancilla needs for the QFT if it is run at the
%speed of data.}
\label{fig:ancNeedsQFT}
\end{minipage}%
\vspace{-0.2in}
\caption{Encoded zero ancilla needs for the QRCA (left), QCLA (middle)
and QFT (right) to run at the speed of data.}
\label{fig:ancNeedsAll}
\end{figure*}

%% The critical path through a 32-bit Quantum Ripple-Carry Adder consists
%% of 1 Z prepare, 63 Hadamards, 32 P gates, 158 T gates, 223 T$^\dagger$
%% gates and 574 CNOTs.  The critical path through a 32-bit Quantum
%% Carry-Lookahead Adder consists of 12 Hadamards, 3 P gates, 21 T gates,
%% 30 T$^\dagger$ gates and 65 CNOTs.  The critical path through a 32-bit
%% Quantum Fourier Transform consists of 1303 Hadamards, 580 T gates, 632
%% T$^\dagger$ gates and 61 CNOTs.

Table~\ref{table:qeclatenciesbyadder} shows for each benchmark the
latency of the critical path in the absence of movement (Column 2),
%using the ion trap latencies from Table~\ref{table:ion_trap_gate_stats},
as well as latencies for the data-dependent and data-independent
(Columns 3 and 4) portions of QEC steps, assuming a QEC operation must
be performed after each useful gate.
% (as in Figure~\ref{fig:adding_qec}) and using the value for
%$t_{qec}$ calculated in Equation~\ref{eq:tqec}.
The minimal running time is the sum of Columns 2 and 3, since these
involve data qubits.  Column 4 corresponds to encoded ancilla
generation time.  Clearly, there is much to be gained in overall
execution time by taking ancilla preparation off the critical path.

%%  & Average Bandwidth Needed \\
%%  & For Non-Transversal Gates \\

% One-column version of the BW needed table
\begin{table}
\small
\begin{center}
\begin{tabular}{|l|c|c|}
\hline
 & Avg Zero Ancilla & Avg $\pi/8$ Ancilla \\
 & Bandwidth Needed & Bandwidth Needed \\
Circuit & For QEC & For $\pi/8$ Gates \\
\hline \hline
32-Bit QRCA & \qrcaAvgBWNeedForQECLogical & \qrcaAvgBWNeedForNonTransLogical \\
32-Bit QCLA & \qclaAvgBWNeedForQECLogical & \qclaAvgBWNeedForNonTransLogical \\
32-Bit QFT & \qftAvgBWNeedForQECLogical & \qftAvgBWNeedForNonTransLogical \\
\hline
\end{tabular}
\end{center}
\vspace{-0.2in}
\caption{Average encoded ancilla bandwidths needed for QEC and
non-transversal gates (in {\it encoded ancillae per millisecond}) if
each circuit is to be executed at the speed of data.}
%% \caption{The average bandwidth of encoded $\pi/8$ ancillae needed for
%% non-transversal one-qubit gates if each circuit is to be executed at
%% the speed of data.}
\label{table:averageBWNeededForQEC}
\end{table}

% Two-column version of the BW needed table
%% \begin{table*}
%% \small
%% \begin{center}
%% \begin{tabular}{|l|c|c|}
%% \hline
%%  & Avg Bandwidth Needed For QEC & Avg Bandwidth Needed for Non-Transversal Gates\\
%% Circuit & (Encoded Zero Ancillae / ms) & (Encoded $\pi/8$ Ancillae / ms) \\
%% \hline \hline
%% 32-Bit QRCA & \qrcaAvgBWNeedForQECLogical & \qrcaAvgBWNeedForNonTransLogical \\
%% 32-Bit QCLA & \qclaAvgBWNeedForQECLogical & \qclaAvgBWNeedForNonTransLogical \\
%% 32-Bit QFT & \qftAvgBWNeedForQECLogical & \qftAvgBWNeedForNonTransLogical \\
%% \hline
%% \end{tabular}
%% \end{center}
%% \vspace{-0.2in}
%% \caption{Average bandwidth of encoded zero ancillae needed for QEC and
%% average bandwidth of encoded $\pi/8$ ancillae needed for
%% non-transversal one-qubit gates if each circuit is to be executed at
%% the speed of data.  Note that bandwidth is given per {\it
%% milli}second.}
%% %% \caption{The average bandwidth of encoded $\pi/8$ ancillae needed for
%% %% non-transversal one-qubit gates if each circuit is to be executed at
%% %% the speed of data.}
%% \label{table:averageBWNeededForQEC}
%% \end{table*}

%We next take a look at the encoded ancilla bandwidth necessary to keep
%each circuit running at the speed of data.
%Figures~\ref{fig:ancNeedsQRCA}, \ref{fig:ancNeedsQCLA} and
%\ref{fig:ancNeedsQFT} show for the QRCA, QCLA and QFT, respectively,
Figure~\ref{fig:ancNeedsAll} shows for the QRCA (left), QCLA (middle)
and QFT (right) the number of encoded ancillae used for QEC which need
to be in the system as execution progresses in order to keep the
circuit operating at the speed of data.  This means that adequate
hardware resources exist to generate and distribute the needed
ancillae in time, but the interaction with data during each QEC step
is still on the critical path of execution.
Table~\ref{table:averageBWNeededForQEC} summarizes this figure by
giving the average encoded ancilla bandwidth necessary for each.

\begin{figure*}
\begin{minipage}{0.33\hsize}
\begin{center}
\epsfig{file=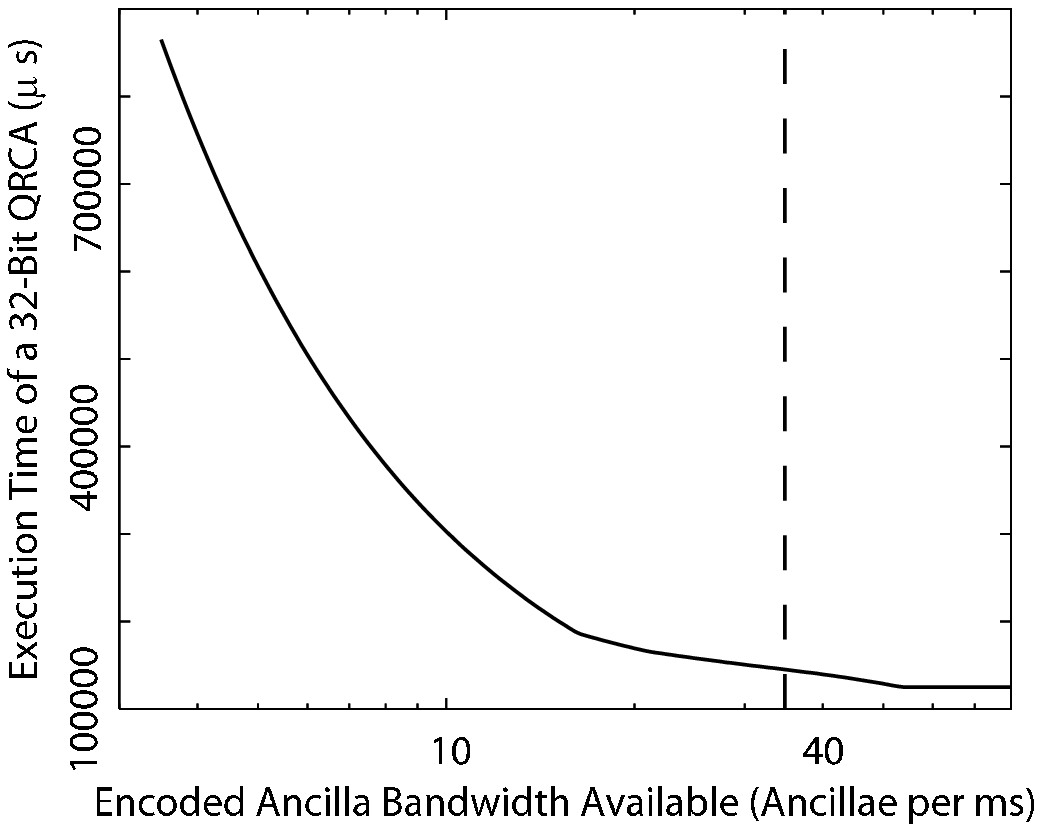,width=\hsize}
\end{center}
%\vspace{-0.2in}
%\caption{The execution time of the QRCA as a function of a steady
%throughput of encoded ancillae.  The vertical line shows the average
%bandwidth from Table~\ref{table:averageBWNeededForQEC}.}
\label{fig:limitedBWAvail32BitRipple}
\end{minipage}%
\hfill%
\begin{minipage}{0.33\hsize}
\begin{center}
\epsfig{file=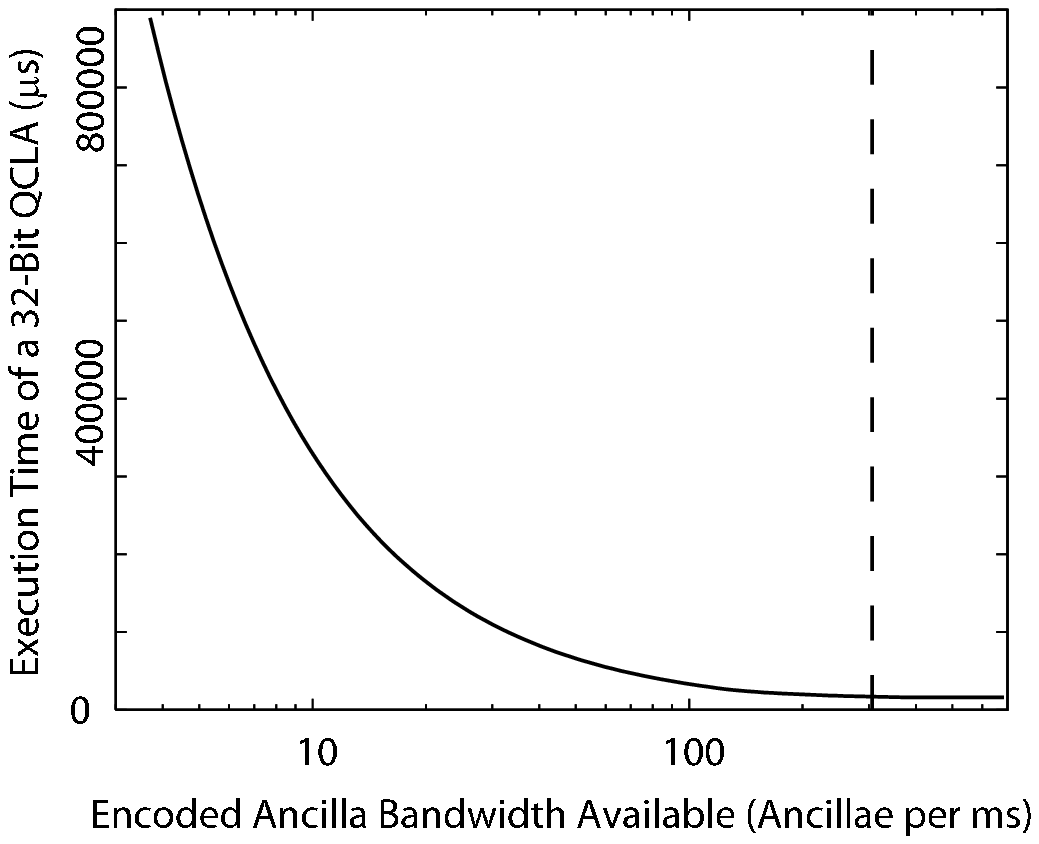,width=\hsize}
\end{center}
%\vspace{-0.2in}
%\caption{The execution time of the QCLA as a function of a steady
%throughput of encoded ancillae.  The vertical line shows the average
%bandwidth from Table~\ref{table:averageBWNeededForQEC}.}
\label{fig:limitedBWAvail32BitCLA}
\end{minipage}%
\hfill%
\begin{minipage}{0.33\hsize}
\begin{center}
\epsfig{file=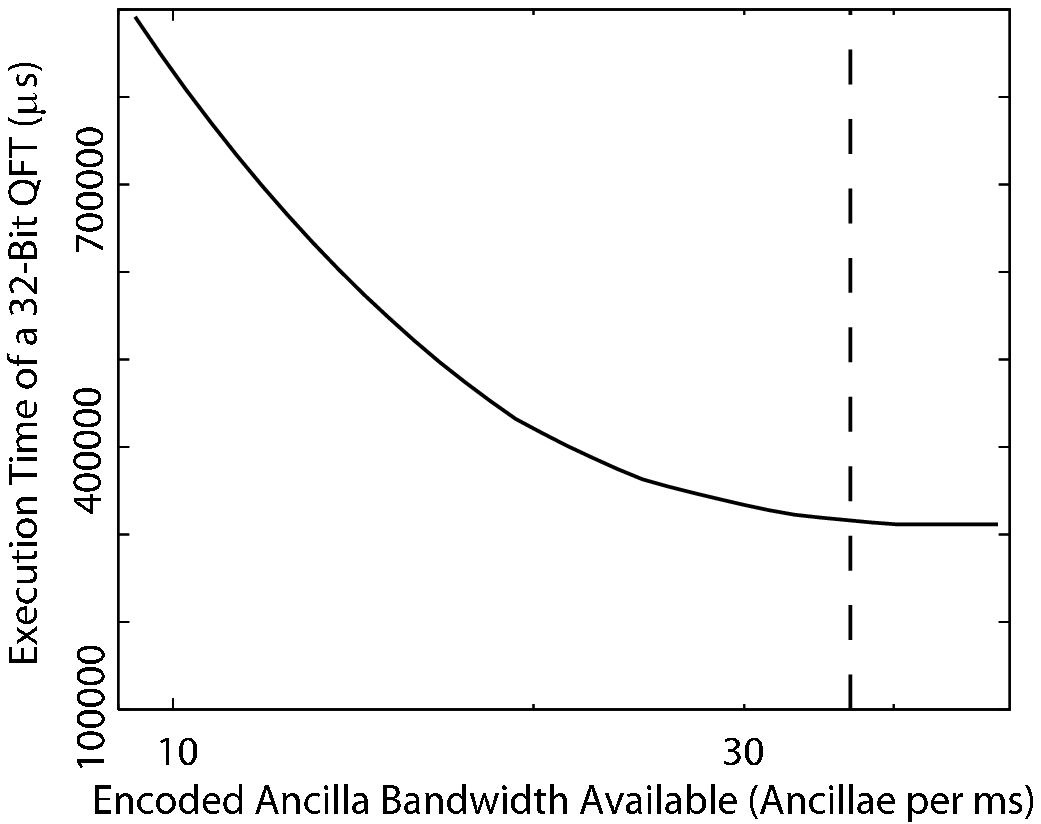,width=\hsize}
\end{center}
%\vspace{-0.2in}
%\caption{The execution time of the QFT as a function of a steady
%throughput of encoded ancillae.  The vertical line shows the average
%bandwidth from Table~\ref{table:averageBWNeededForQEC}.}
\label{fig:limitedBWAvail32BitQFT}
\end{minipage}%
\vspace{-0.2in}
\caption{The execution time of the QRCA (left), QCLA (middle) and QFT
(right) as a function of a steady throughput of encoded zero ancillae.
The vertical line in each shows the average bandwidth for that circuit
from Table~\ref{table:averageBWNeededForQEC}.}
\label{fig:limitedBWAvail32BitAll}
\end{figure*}

%% \begin{figure}
%% \begin{center}
%% \epsfig{file=graphs/limitedBWAvail32BitRippleZeroLat.eps,width=0.9\hsize}
%% \end{center}
%% \vspace{-0.2in}
%% \caption{The execution time of the QRCA as a function of a steady
%% throughput of encoded ancillae.  The vertical line shows the average
%% bandwidth from Table~\ref{table:averageBWNeededForQEC}.}
%% \label{fig:limitedBWAvail32BitRipple}
%% \end{figure}

%% \begin{figure}
%% \begin{center}
%% \epsfig{file=graphs/limitedBWAvail32BitCLAZeroLat.eps,width=0.9\hsize}
%% \end{center}
%% \vspace{-0.2in}
%% \caption{The execution time of the QCLA as a function of a steady
%% throughput of encoded ancillae.  The vertical line shows the average
%% bandwidth from Table~\ref{table:averageBWNeededForQEC}.}
%% \label{fig:limitedBWAvail32BitCLA}
%% \end{figure}

%% \begin{figure}
%% \begin{center}
%% \epsfig{file=graphs/limitedBWAvail32BitQFTZeroLat.eps,width=0.9\hsize}
%% \end{center}
%% \vspace{-0.2in}
%% \caption{The execution time of the QFT as a function of a steady
%% throughput of encoded ancillae.  The vertical line shows the average
%% bandwidth from Table~\ref{table:averageBWNeededForQEC}.}
%% \label{fig:limitedBWAvail32BitQFT}
%% \end{figure}

These averages do not take into account the handling of peak periods.
In reality, the encoded ancilla bandwidth necessary to run a circuit
optimally may be higher than the average bandwidth.
%Figures~\ref{fig:limitedBWAvail32BitRipple},
%\ref{fig:limitedBWAvail32BitCLA}, and \ref{fig:limitedBWAvail32BitQFT}
%show for the QRCA, QCLA and QFT, respectively, the circuit execution
Figure~\ref{fig:limitedBWAvail32BitAll} shows for the QRCA (left),
QCLA (middle) and QFT (right) the circuit execution time assuming a
steady throughput of encoded ancillae being generated, as specified on
the x-axis.
%  For all three circuits, the average bandwidth
%results in near-optimal execution time.
%  To keep the number of encoded ancillae in
%the system from escalating without limit, we limit the total number of
%encoded ancillae in the system to the maximum number needed at one
%time for each circuit, which is 18 for the QRCA, 138 for the QCLA and
%12 for the QFT.
These graphs show us the sustained ancilla bandwidth necessary to run
each circuit at near-optimal speed, but these are only estimates since
they lack the details of movement and layout.  In
Section~\ref{sec:factory}, we examine the associated hardware needs.
% for generating these ancilla bandwidths.
%, followed by an investigation of the interconnect
%distribution network required to support all this communication.

\subsection{Non-Transversal One-Qubit Gates}\label{sec:nonTransBW}

The encoded ancilla bandwidth needs discussed in
Section~\ref{sec:circuit_needs} for our three benchmarks include only
zero ancillae needed for error correction.  Non-transversal one-qubit
gates account for \qrcaOneQNonTransGatePercent\%,
\qclaOneQNonTransGatePercent\% and \qftOneQNonTransGatePercent\% of
our QRCA, QCLA and QFT benchmarks circuits, respectively, when using
the [[7,1,3]] encoding.  As explained in Section~\ref{sec:pi8circuit},
the execution of a non-transversal encoded gate is performed with the
use of a $\pi/8$ encoded ancilla qubit.  Column 3 in
Table~\ref{table:averageBWNeededForQEC} shows the corresponding
$\pi/8$ ancilla bandwidth needed for each benchmark to achieve a
runtime limited only by the speed of data (the sum of Columns 2 and 3
in Table~\ref{table:qeclatenciesbyadder}).

\section{Ancilla Factory Layout}\label{sec:factory}

%% \begin{itemize}
%% \item argument: want ancilla factories
%% \item FIG and argument: classically, pipeline; why pipelining doesn't help (synchronous, no specialization)
%% \item TABLE: throughput vs. area for one-channel vs. two-channel (need multiple codes or automate this)
%% \item TABLE: preparation with non-transversal gate: before vs. after encoding (need multiple code or automate this)
%% \end{itemize}

In this section, we shall explore the design space of possible ancilla
factories and determine the hardware resources necessary to produce
encoded ancillae at the bandwidths calculated in
Sections~\ref{sec:circuit_needs} and \ref{sec:nonTransBW} in order to
take ancilla generation off the critical path of execution.

\vspace{-11pt}
\subsection{Ion Trap Abstraction}

%% \begin{itemize}
%% \item movement
%% \item gates
%% \item macroblocks: area will be computed in terms of these
%% \end{itemize}
Our area calculations are done using an abstraction of ion trap
technology~\cite{seidelin06}, described here.
%We will perform area calculations using an abstraction of the ion trap
%technology~\cite{???}, described here.
\vspace{-0.1in}
\paragraph{Qubits:} 
A single qubit capable of holding one bit of quantum state is an ion.  
%A single qubit which holds one bit of quantum state is an ion.  
The physical implementation of a qubit is actually more complicated,
but for our purposes, we may represent each qubit as a single
ion. \vspace*{-11pt}
%consider each qubit to be a single charged particle.
%\vspace{-0.2in}
\paragraph{Movement:} Electrodes are used to create
potential wells in which qubits (ions) are trapped.  Potential wells
and the ions within are moved via an application of precise pulse
sequences to the electrodes.  Moving an ion around a corner takes more
time than moving straight~\cite{hensinger2006tji}.  The latency
numbers we use are shown Table~\ref{table:ion_trap_move_stats}.

%% In ion traps, qubit speed when turning a corner may be slower than
%% during straight channel movement.
%%  shows the latency numbers we
%% use for qubit movement.  Straight Move latency is the time
%% to move between adjacent macroblocks to and the Turn time is
%% the time for a qubit to turn a corner.  As with gates, we will
%% keep our calculations symbolic when possible to allow for changes in
%% move latenices.

\begin{table}
\small
\begin{center}
\begin{tabular}{|l|c|c|}
\hline
%Physical & Latency & Latency \\
%Operation & Symbol & ($\mu$s) \\
Physical Operation & Latency Symbol & Latency ($\mu$s)\\
\hline \hline
Straight Move & $t_{move}$ & \tMoveLat \\
Turn & $t_{turn}$ & \tTurnLat \\
\hline
\end{tabular}
\end{center}
\vspace{-0.2in}
\caption{Latency values for the two types of move operations in ion
trap technology~\cite{steane2004bbg,thaker2006qmh,isailovic2006ins}.
A Straight Move is across a single macroblock
(Figure~\ref{fig:macroblocks}).}
\label{table:ion_trap_move_stats}
\end{table}

%By applying precise pulse sequences to these electrodes, the
%potential wells are moved, thus moving the trapped ions as well.
%Through application of precise pulse sequences to these electrodes, the
%potential wells may be moved, thus moving the trapped ions as well.
%\vspace{-0.4in}
\paragraph{Gates:} \vspace*{-11pt} A gate is performed by firing precise laser pulses
at a trapped ion.
%  Alternatively, it has been proposed that a gate may
%be more easily performed by holding steady laser beams and moving ions
%through them~\cite{leibfried2007tql}.  In either case,
We may abstract away the physics and consider that a gate is performed
by arrival at certain special ``gate locations'' in the layout. \vspace*{-11pt}

\begin{figure}
%\vspace{0.1in}
\begin{center}
\epsfig{file=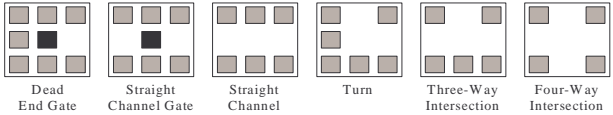,width=\hsize}
\end{center}
\vspace{-0.2in}
\caption{The abstract building blocks of our layouts.  Black boxes are
gate locations (which may not occur in an intersection), grey boxes
are abstract ``electrodes,'' and wide white channels are valid
paths for qubit movement.}
\label{fig:macroblocks}
%\vspace{-0.1in}
\end{figure}

%\vspace{-0.2in}
\paragraph{Macroblocks:} Since qubit movement is performed by
electrodes whose position is fixed at fab time, certain ``channels''
for qubit movement are also set at fab time.  The details of electrode
structure are still evolving, so determining area in terms of number
of ion traps is a bit ambiguous.  For this reason, we use the {\it
macroblocks} shown in Figure~\ref{fig:macroblocks} as the basic
building blocks of our layouts.  Each macroblock has one or more
``ports'' through which qubits may enter and exit and which connect to
an adjacent macroblock.  To perform a gate operation, all involved
qubits must enter a valid gate location (a black square in our
macroblocks) and remain there for the duration of the gate.  Our area
numbers are all calculated in terms of macroblock count.

\subsection{Data Qubit Area}

%% \begin{itemize}
%% \item Figure~\ref{fig:data_area}: room for two encoded qubits
%% \item everything else is interconnect or ancilla gen
%% \item discuss transversal gates: encoded T ancilla prepared, so
%% interaction with data is still transversal
%% \end{itemize}

%We assume that data in a compute region is encoded using a scheme
%which allows transversal two-qubit gates, such as the Level 1 Steane
%[[7,1,3]] encoding methodology \cite{???} which we'll be using in our
%calculations.
Over the run of a quantum circuit, encoded data must perform four
distinct types of operations: transversal one-qubit gates,
non-transversal one-qubit gates, transversal two-qubit gates and QEC
steps.  As described in Section~\ref{sec:pi8circuit}, a
non-transversal one-qubit gate may be performed by preparing a
specific encoded ancilla and interacting it transversally with the
data qubit.  Likewise, the data/ancilla interaction portion of a QEC
step involves a transversal two-qubit gate.
%In both cases, there may be a one-qubit gate correction applied to the
%encoded data at the end.
In the end, the main operations the encoded data must support are
transversal one- and two-qubit gates.
%This means that encoded data qubits essentially need to
%be able to perform transversal one- and two-qubit gates (numbers 1 and
%3 above).

\begin{figure}
%\begin{minipage}{.3\hsize}
\begin{center}
\epsfig{file=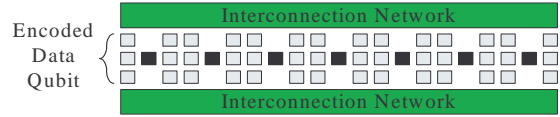,width=.9\hsize}
\end{center}
\vspace{-0.2in}
\caption{Layout of a single encoded data qubit.}
%  Each row contains enough
%room for one encoded data qubit; the rows are connected to allow
%application of a transversal two-qubit gate, and each qubit may access
%the interconnect network on its side.}
\label{fig:data_area}
%\end{minipage}
%\hfill
%\begin{minipage}{.65\hsize}
% fig:anc_fac_line used to be here
%\end{minipage}
\end{figure}

To support these major operations, we use single compute regions as
shown in Figure~\ref{fig:data_area}.  The design consists of a single
column of Straight Channel Gate Macroblocks with enough room for a
single encoded qubit (seven macroblocks for the [[7,1,3]] CSS code),
with access on either side to whatever interconnect network is being
used.
%  This allows for the operation
%of both one- and two-qubit transversal gates.  Each encoded qubit may
%access the interconnect on its side of the compute region, whether for
%ballistic movement or interaction with EPR qubits for teleportation.
Thus, if we are encoding each qubit into $m$ physical qubits, the
total area used by data is $m \times n_q$,
%\begin{eqnarray}
%Total\ Data\ Area = m \times n_q
%\end{eqnarray}
where $n_q$ is the total number of data qubits (including data
ancillae) in the circuit.
%  Some area might be saved by having special
%compute regions for one-qubit gates, however the circuit execution
%time is dominated by two-qubit gates (including those used for
%non-transversal one-qubit gates and for QEC steps), so we'll use this
%layout for simplicity.

\subsection{Simple Ancilla Factories}\label{sec:simple_fac}

%% \begin{itemize}
%% \item FIG (Figure~\ref{fig:anc_prep}): prepare, FT and verify,
%% throughput
%% \item ??? show prep and verify circuit, or just reference
%% \item FIG (Figure~\ref{fig:line_1col}): first attempt: prior work:
%% line (cite CQLA): 7 + 3 for verification
%% \item EQUATION: computer area and throughput for line
%% \item TABLE (Table~\ref{table:matchedbw}): for each adder, match BW by
%% using multiple lines; metric: area
%% \item Figure~\ref{fig:adder_microarch2}: microarchitecture drawn to
%% scale
%% \item Figure\ref{fig:graph3}: area needed for ancilla gen vs. total
%% execution time: combine with earlier BW-limiting graph
%% \end{itemize}

\begin{figure}
\begin{center}
\epsfig{file=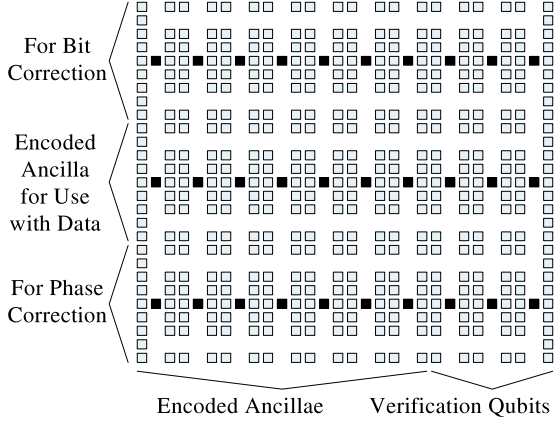,width=.9\hsize}
\end{center}
\vspace{-0.2in}
\caption{An ancilla factory for the circuit in
  Figure~\ref{fig:anc_prep_options}c.  Each row of gates
  %and its two adjacent communication rows 
  generates and verifies one of
  the three encoded zero ancillae, then bit and phase correction are
  performed.}
%% the [[7,1,3]] code based on
%% the layout in QLA~\cite{???}.  Stateless physical qubits enter from
%% the top, perform the circuit in Figure~\ref{???} using the seven gate
%% locations, then exit the bottom as an encoded ancilla. (b) A
%% reduced-area version of the layout, taking advantage of the fact that
%% only four gate locations are needed for the CNOT portion of the
%% circuit.  This one requires that the physical prepares and Hadamards
%% have been completed.}
\label{fig:anc_fac_line}
\end{figure}

We now focus on designing an {\it ancilla factory}, a concept first
proposed in~\cite{steane1999stp}.  An ancilla factory is a portion of
the layout which consumes stateless physical ancillae and produces a
steady stream of encoded ancillae at some rate.
%We begin by addressing the generation of zero ancillae and
%follow with an ancilla factory for generating encoded $\pi/8$ ancillae
%in Section~\ref{sec:pi8factory}.
%We first target the generation of zero ancillae, then we tackle
%an ancilla factory for generating encoded $\pi/8$ ancillae in Section~\ref{sec:pi8factory}.
Figure~\ref{fig:anc_fac_line} shows a simple ancilla factory to
execute the circuit in Figure~\ref{fig:anc_prep_options}c.  Each row
of gates has room for ten physical qubits, seven to be encoded
and three for verification.
%seven physical qubits to be encoded plus three more for verification.  
The adjacent rows are used for communicating.
When all three are encoded and verified, the middle one is
bit-corrected by the top one and phase-corrected by the bottom one.
%%     The circuit in Figure~\ref{fig:anc_prep_options}c
%% requires three zero ancillae to be encoded and verified.    Seven junk qubits enter from the top, use the seven
%% gate locations to perform the circuit in Figure~\ref{???}, then exit
%% out the bottom as a single encoded zero ancilla.
Using a hand-optimized schedule, the total latency of a single ancilla
preparation is approximately: $t_{prep} + 2 \times t_{meas} + 6 \times t_{2q} + 2 \times t_{1q} + 8
\times t_{turn} + 30 \times t_{move}$.

Substituting in the ion trap latencies in
Tables~\ref{table:ion_trap_gate_stats} and
\ref{table:ion_trap_move_stats}, the layout in
Figure~\ref{fig:anc_fac_line} has a total latency of \ancFacSimpleLat
$\mu$s with a throughput of
%  Thus, the layout in Figure~\ref{fig:anc_fac_line} has a throughput of
\ancFacSimpleBWLogical\ encoded ancillae per millisecond and an area of
90 macroblocks.  Using this simple ancilla factory, we could produce
any desired bandwidth of encoded ancillae by replicating the layout as
many times as necessary.
%simply duplicating this layout as many times as necessary.
Unfortunately this design is inefficient in that the verification
qubits needlessly take up space during the seven-qubit zero encoding
procedure.  To combat this inefficiency we instead consider a
pipelined approach.

%% It is evident, however, that there is a good deal of inefficiency in
%% this design, such as the fact that the verification qubits are
%% needlessly taking up space during the seven-qubit zero encoding
%% procedure.  As an alternate approach, we observe that our ancilla
%% factories are performing the same circuit repeatedly, and so we use an
%% appropriate technique from classical design: pipelining.

\subsection{Pipelined Ancilla Factories}\label{sec:pipe_fac}

%% \begin{figure}
%% \begin{center}
%% \epsfig{file=figures/pipe_anc_prep_2col2.eps,width=.8\hsize}
%% \end{center}
%% \vspace{-0.2in}
%% \caption{(a) A simple layout for performing the CNOT portion of the
%% circuit in Figure~\ref{fig:trans_vs_nontrans}b, with three CNOTs
%% performed in parallel at a time.  (b) A pipelined version of the
%% layout in (a).}
%% \label{fig:pipe_anc_prep_2col}
%% \end{figure}
Classically, pipelining a circuit is done by inserting synchronization
points (registers) into the circuit's datapath to enable logic reuse,
thereby increasing throughput with a small increase in latency.  We
can apply a similar technique to our ancilla factory layout in an
effort to improve area utilization. Due to the precise electrode and
laser pulse sequences needed to implement movement and gates, ion trap
layouts are by definition synchronous without additional
synchronization elements.  Instead, we must add a set of communication
channels between pipeline stages allowing qubit movement for maximum
gate location occupancy.

\begin{figure}
\begin{center}
\epsfig{file=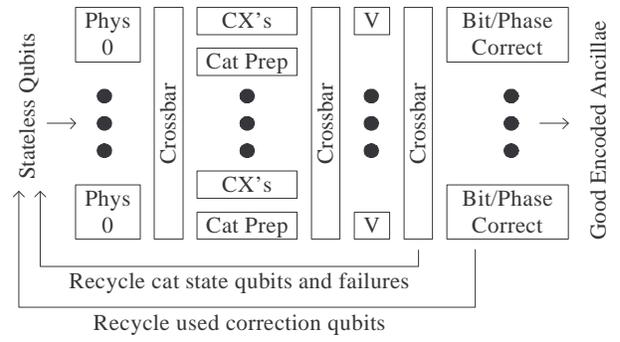,width=\hsize}
\end{center}
\vspace{-0.2in}
\caption{A fully pipelined encoded zero ancilla creation unit
implementing the circuit in Figure~\ref{fig:anc_prep_options}c.}
\label{fig:pipe_full_prep_2col}
%\end{minipage}
\end{figure}
%\hfill
%\begin{minipage}{0.45\hsize}
\begin{figure}
\begin{center}
\epsfig{file=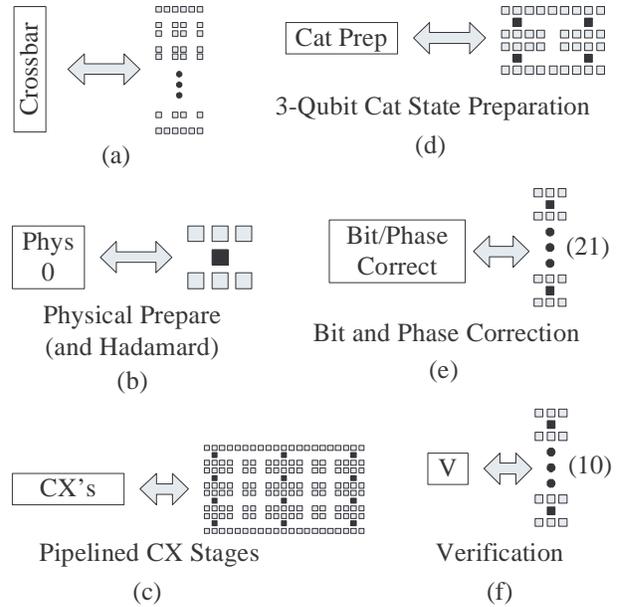,width=\hsize}
\end{center}
\vspace{-0.2in}
\caption{A layout of each unit in
Figure~\ref{fig:pipe_full_prep_2col}.}
\label{fig:pipe_full_prep_2col_exp}
\end{figure}
%\end{minipage}
%\end{figure*}

%% This is not to say, however, that there are no other benefits to
%% pipelining.  One of the primary benefits is the input and output
%% ``ports'' inherent in the pipeline.  This allows us to focus the
%% generation point of freshly encoded zero ancillae near data qubits,
%% as we discuss further in Section~\ref{???}.  But first, we
%% investigate the characteristics of a pipelined ancilla factory.

\subsubsection{Encoded Zero Ancilla Factory}
We consider the entire circuit for fault tolerant encoded zero ancilla
creation (Figure~\ref{fig:anc_prep_options}c).
Figure~\ref{fig:pipe_full_prep_2col} shows a fully pipelined
microarchitecture for this circuit, which consists of four
stages.  Each stage contains a number of functional units for its
subcircuit such that the output bandwidth of one stage is matched to
the input bandwidth of the next.  Adjacent stages are
separated by a crossbar (Figure~\ref{fig:pipe_full_prep_2col_exp}a),
which consists of two vertical columns, fully connected horizontally,
one for upwards movement, the other for downwards.
%  In this way, each qubit may move from one stage to the next without
%worrying about congestion, since qubits moving in opposite directions
%won't impede each other.

Stage 1 consists of preparing a junk physical qubit into the zero
state with an optional Hadamard gate at a single gate location
(Figure~\ref{fig:pipe_full_prep_2col_exp}b).  Even though only some of these
qubits need the Hadamard, we group them all into the same set of
functional units.

\begin{table*}
\small
\begin{center}
\begin{tabular}{|l|c||c|c|c|c|c|}
\hline
 & & Latency & & \multicolumn{2}{c|}{BW (qubits/ms)} & Area \\
Functional Unit & Symbolic Latency & ($\mu$s) & Stages & In & Out & \\
\hline \hline
Zero Prep & $t_{prep} +t_{1q} + 2 \times t_{turn} + t_{move}$ & \stageOneLat & 1 & \stageOneInBWPhysical & \stageOneOutBWPhysical & \stageOneArea \\
CX Stage & $3 \times t_{2q} + 6 \times t_{turn} + 5 \times t_{move}$ & \stageTwoCNOTLat & 3 & \stageTwoCNOTInBWPhysical & \stageTwoCNOTOutBWPhysical & \stageTwoCNOTArea \\
Cat State Prep & $2 \times t_{2q} + 4 \times t_{turn} + 2 \times t_{move}$ & \stageTwoCatLat & 2 & \stageTwoCatInBWPhysical & \stageTwoCatOutBWPhysical & \stageTwoCatArea \\
Verification & $t_{meas} + t_{2q} + 2 \times t_{turn} + 2 \times t_{move}$ & \stageThreeLat & 1 & \stageThreeInBWPhysical & \stageThreeOutBWPhysical & \stageThreeArea \\
B/P Correction & $t_{meas}\!+\!2\!\times\!t_{2q}\!+\!6\!\times\!t_{turn}\!+\!8\!\times\!t_{move}$ & \stageFourLat & 1 & \stageFourInBWPhysical & \stageFourOutBWPhysical & \stageFourArea \\
\hline
\end{tabular}
\end{center}
\vspace{-0.2in}
\caption{For each functional unit in
Figure~\ref{fig:pipe_full_prep_2col}, Column 2 gives its symbolic
latency.
% including two turns for moving between stages, but no
%straight moves.
The remaining columns give numeric values using our
ion trap assumptions.  ``Stages'' is the number of pipeline stages
within the functional unit itself, and ``Area'' is given in number of
macroblocks.}
\label{table:func_units}
\end{table*}

%Abbreviated table
\begin{table}
%\begin{minipage}{0.49\hsize}
\small
\begin{center}
\begin{tabular}{|l|c|c|c|}
\hline
 & Unit & Total & Total \\
Functional Unit & Count & Height & Area \\
\hline \hline
Zero Prepare & \stageOneCount & \stageOneTotalHeight & \stageOneTotalArea \\
CX Stage & \stageTwoCNOTCount & \stageTwoCNOTTotalHeight & \stageTwoCNOTTotalArea \\
Cat State Prepare & \stageTwoCatCount & \stageTwoCatTotalHeight & \stageTwoCatTotalArea \\
Verification & \stageThreeCount & \stageThreeTotalHeight & \stageThreeTotalArea \\
B/P Correction & \stageFourCount & \stageFourTotalHeight & \stageFourTotalArea \\
\hline
\end{tabular}
\end{center}
\vspace{-0.2in}
\caption{The functional unit counts and stage characteristics for the
encoded zero ancilla factory in Figure~\ref{fig:pipe_full_prep_2col}.
The CX and Cat Prepare units in Stage 2 are bandwidth matched to a
ratio of 7 to 3 (which is appropriate for verification), and then the
other stages are matched to this.}
\label{table:func_unit_counts}
%\end{minipage}
\end{table}

Stage 2 consists of two types of units.  Looking at the CX portion
of the ancilla prepare circuit in Figure~\ref{fig:trans_vs_nontrans}b,
we see that the first three CX's can be performed in parallel, as can
the next three, followed by the final three.  Thus, we may use the
pipelined layout in Figure~\ref{fig:pipe_full_prep_2col_exp}c for this
functional unit, with three sets of qubits (each performing three
CX's with one idle qubit) in this functional unit at a time.  The Cat
Prep units (Figure~\ref{fig:pipe_full_prep_2col_exp}d) create a
three-qubit cat state out of three physical zero ancillae by
performing two CX's in succession.
%Thus, we use the pipelined circuit depicted in

Verification of the encoded zero ancillae using the cat states is
performed in Stage 3 and involves performing three CX's in
parallel and then measuring the cat state qubits to determine success
or failure of the encoded ancilla.  Since the encoded ancilla
qubits must wait for the measurement to complete, we need 10
macroblocks, one for each qubit as shown in
Figure~\ref{fig:pipe_full_prep_2col_exp}e.  When this is done, the three
qubits of the cat state are recycled immediately, as well as the other
seven qubits if the verification failed.

Finally, in Stage 4, a verified encoded zero ancilla A is
first bit-corrected by a verified encoded zero ancilla B
and then phase-corrected by a verified encoded zero
ancilla C.  Since we need storage room for A plus room to measure both
B and C in parallel (allowing us to overlap these measurements in
time), each such functional unit needs space for three encoded
ancillae, as shown in Figure~\ref{fig:pipe_full_prep_2col_exp}f.

%Full table
%% \begin{table*}
%% \small
%% \begin{center}
%% \begin{tabular}{|l|c|c|c|c|c|}
%% \hline
%%  & & Total Height & Total Input BW & Total Output BW & Total Area \\
%% Functional Unit & Count & (in macroblocks) & (qubits/ms) & (qubits/ms) & (in macroblocks) \\
%% \hline \hline
%% Zero Prepare & \stageOneCount & \stageOneTotalHeight & \stageOneTotalInBWPhysical & \stageOneTotalOutBWPhysical & \stageOneTotalArea \\
%% CX Stage & \stageTwoCNOTCount & \stageTwoCNOTTotalHeight & \stageTwoCNOTTotalInBWPhysical & \stageTwoCNOTTotalOutBWPhysical & \stageTwoCNOTTotalArea \\
%% Cat State Prepare & \stageTwoCatCount & \stageTwoCatTotalHeight & \stageTwoCatTotalInBWPhysical & \stageTwoCatTotalOutBWPhysical & \stageTwoCatTotalArea \\
%% Verification & \stageThreeCount & \stageThreeTotalHeight & \stageThreeTotalInBWPhysical & \stageThreeTotalOutBWPhysical & \stageThreeTotalArea \\
%% B/P Correction & \stageFourCount & \stageFourTotalHeight & \stageFourTotalInBWPhysical & \stageFourTotalOutBWPhysical & \stageFourTotalArea \\
%% \hline
%% \end{tabular}
%% \end{center}
%% \vspace{-0.2in}
%% \caption{The functional unit counts and stage characteristics for the
%% encoded zero ancilla factory in Figure~\ref{fig:pipe_full_prep_2col}.
%% The CX and Cat Prepare units in Stage 2 are bandwidth matched to a
%% ratio of 7 to 3 (which is appropriate for verification), and then the
%% other stages are matched to this.}
%% \label{table:func_unit_counts}
%% \end{table*}

% Note: The two tables for the zero ancilla factory are firest
% referenced here.

Table~\ref{table:func_units} summarizes the latency breakdown for
each stage of the pipeline and provides numerical values for
various characteristics of each functional unit under our ion trap
assumptions.
%All of this is summarized in Table~\ref{table:func_units}.  For each
%of the functional units, the second column gives the symbolic latency
%using basic latency parameters.
%%  The {\it turn} and {\it move} counts
%%are specific to the macroblocks in Figure~\ref{fig:macroblocks},
%%however the concepts of ``turn,'' ``intersection'' and ``straight
%%channel'' are general enough that the same or a similar set of
%%macroblocks could be used for other technologies.
%The right five columns give numerical values for various
%characteristics of each functional unit under our ion trap
%assumptions.  
Note that Stages 3 and 4 have input bandwidth different from output
bandwidth due to the fact
%\vspace{-2.4in}
%\noindent 
that some qubits are used up and recycled in
these stages.
%Using these numbers, we determine functional unit
%counts that (approximately) match throughput through the stages.  A
%chief goal in our ancilla factory is near full utilization of
%resources, meaning that we want to match bandwidths to avoid unit
%idleness.
To achieve high resource utilization, we determine unit count by
matching bandwidth between successive stages.  The results are shown
Table~\ref{table:func_unit_counts}.

For the crossbars, we use a two-column design, one column for upwards
movement, the other for downwards, in order to avoid congestion.
However, physical qubits exiting Stage 1 are funneled inward to the
much smaller Stage 2, so we use a single column crossbar since
bi-directionality is likely unnecessary.  The total crossbar area is
thus \stageOneTotalHeight\ + 2 * \stageThreeTotalHeight\ + 2 *
\stageFourTotalHeight\ = \totalCrossbarArea\ macroblocks, and the
total functional unit area is \stageOneTotalArea\ +
\stageTwoTotalArea\ + \stageThreeTotalArea\ + \stageFourTotalArea\ =
\totalComputeArea\ macroblocks, resulting in a total area of
\totalAncFacArea\ macroblocks.
%\end{eqnarray}

For overall throughput, we take the minimum throughput among the
stages.
%%   When we calculated latency earlier, we ignored straight
%% channel movement in the crossbars because it was hard to estimate in
%% advance.  Given the relatively small size of our factory and the
%% relatively trivial time for a straight channel move, we shall ignore
%% the effect of these moves on stage latency.
The bottleneck in the factory is the CX Stage.  Each seven physical
qubits out of this stage correspond to an encoded zero ancilla.
Approximately 99.8\% of these qubits are successfully verified (using
the results of our Monte Carlo simulations mentioned in
Section~\ref{sec:ftsp}), and two-thirds of them are then used to
correct the other third.  Thus, the overall throughput of our zero
ancilla factory is:
%\begin{eqnarray}
$\frac{\stageTwoCNOTInBWPhysical}{7} \times 0.998 \times \frac{1}{3} = \totalAncFacBWLogical$ encoded ancillae / ms.
%\end{eqnarray}

\subsubsection{Encoded $\pi/8$ Ancilla Factory}\label{sec:pi8factory}

In Section~\ref{sec:nonTransBW}, we showed that a non-trivial supply
of encoded $\pi/8$ ancillae are also needed by our circuits.  The
circuit in Figure~\ref{fig:pi8_anc_prep}b shows how to turn a zero
ancilla generated by our pipelined ancilla factories into an encoded
$\pi/8$ ancilla.
%  We pipeline this circuit as well, not merely for the
%benefit of input/output ports, but also because the different number
%of qubits involved at various points in the circuit means that
%resources would be idle if this were done ``in place.''
This circuit may be divided into four stages: 1) Cat State Prepare, 
2) Transversal Controlled-Z/S/X, plus Transversal $\pi/8$,
3) Decode,
4) One-qubit H, One-qubit Measure, Transversal Z conditional on measurement.
%\end{enumerate}

%% \begin{figure}
%% \begin{center}
%% \epsfig{file=figures/pi8gen_layout.eps,width=0.9\hsize}
%% \end{center}
%% \vspace{-0.2in}
%% \caption{Basic layout blocks of a pipelined encoded $\pi/8$ ancilla
%% factory.}
%% \label{fig:pi8gen_layout}
%% \end{figure}

%% The layout in Figure~\ref{fig:pi8gen_layout}a may be used for the Cat
%% State Prepare and Decode, since at any time five qubits are idle while
%% the other two are performing a gate.  (Note that this is a seven-qubit
%% Cat State Prepare, as opposed to the three-qubit cat state used in
%% encoded zero ancilla preparation, though the circuits are similar.)
%% The layout in Figure~\ref{fig:pi8gen_layout}b may be used for Stages 2
%% and 4, since they are fully transversal, while Stage 3 requires only a
%% single gate location.  We also note that a separate ``holding pen''
%% (essentially Figure~\ref{fig:pi8gen_layout}b, but without the need for
%% gate functionality) is needed during Stage 3 and part of Stage 4 for
%% the bottom qubit in Figure~\ref{fig:pi8_anc_prep}b, since it is doing
%% nothing but must still exist somewhere.

\begin{table*}
\small
\begin{center}
\begin{tabular}{|l|c||c|c|c|c|}
\hline
Stage & Symbolic Latency & Latency & In BW & Out BW & Area \\
\hline \hline
Cat State Prepare & $7 \times t_{2q} + 14 \times t_{turn} + 8 \times t_{move}$ & \piEightCatPrepLat & \piEightCatPrepInBWPhysical & \piEightCatPrepOutBWPhysical & \piEightCatPrepArea \\
Transversal CX/CS/CZ/$\pi/8$ & $3 \times t_{2q} + 2 \times t_{turn} + 3 \times t_{move}$ & \piEightTransXSZLat & \piEightTransXSZInBWPhysical & \piEightTransXSZOutBWPhysical & \piEightTransXSZArea \\
Decode (plus Store) & $7 \times t_{2q} + 14 \times t_{turn} + 8 \times t_{move}$ & \piEightDecodeLat & \piEightDecodeInBWPhysical & \piEightDecodeOutBWPhysical & \piEightDecodeArea \\
H/M/Transversal Z & $t_{meas}\!+\!2\!\times\!t_{1q}\!+\!2\!\times\!t_{turn}\!+\!2\!\times\!t_{move}$ & \piEightHMTransZLat & \piEightHMTransZInBWPhysical & \piEightHMTransZOutBWPhysical & \piEightHMTransZArea \\
\hline
\end{tabular}
\end{center}
\vspace{-0.2in}
\caption{For each stage in the encoded $\pi/8$ ancilla generation
circuit,
%in Figure~\ref{fig:pi8_anc_prep}b, 
we give its symbolic latency, plus numeric values for various
characteristics of the stage under our ion trap assumptions.}
\label{table:pi8_func_units}
\end{table*}

%Full table
%% \begin{table*}
%% \small
%% \begin{center}
%% \begin{tabular}{|l|c|c|c|c|c|}
%% \hline
%% Stage & Unit Count & Total Height & Total In BW & Total Out BW & Total Area \\
%% \hline \hline
%% Cat State Prepare & \piEightCatPrepCount & \piEightCatPrepTotalHeight & \piEightCatPrepTotalInBWPhysical & \piEightCatPrepTotalOutBWPhysical & \piEightCatPrepTotalArea \\
%% Transversal CX/CS/CZ/$\pi/8$ & \piEightTransXSZCount & \piEightTransXSZTotalHeight & \piEightTransXSZTotalInBWPhysical & \piEightTransXSZTotalOutBWPhysical & \piEightTransXSZTotalArea \\
%% Decode (plus Store) & \piEightDecodeCount & \piEightDecodeTotalHeight & \piEightDecodeTotalInBWPhysical & \piEightDecodeTotalOutBWPhysical & \piEightDecodeTo\talArea \\
%% H/M/Transversal Z & \piEightHMTransZCount
%% & \piEightHMTransZTotalHeight &
%% \piEightHMTransZTotalInBWPhysical &
%% \piEightHMTransZTotalOutBWPhysical &
%% \piEightHMTransZTotalArea \\
%% \hline
%% \end{tabular}
%% \end{center}
%% \vspace{-0.2in}
%% \caption{The functional unit counts and characteristics for each stage
%% of our final $\pi/8$ ancilla factory.}
%% \label{table:pi8_func_unit_counts}
%% \end{table*}

Table~\ref{table:pi8_func_units} shows the characteristics of each of
these stages.  Note that bandwidths here are in physical qubits, which
is why Stages 1 and 3 have differing bandwidths despite having the
same latency.  We now match bandwidths just as we did for the zero
ancilla factory in order to get close to full utilization.
Table~\ref{table:pi8_func_unit_counts} shows the the final unit counts
of our $\pi/8$ ancilla factory.  Note that only half the qubits
consumed by Stage 2 come from Stage 1 (the others come from an encoded
zero ancilla factory).
% so the necessary output bandwidth of Stage 1 is halved.

The total stage heights are different enough that an exact layout
would likely require partially folding some stages into others and
simulating execution to determine exact crossbar sizes needed to avoid
congestion.  For our purposes, we will allocate two columns to each
crossbar, since qubits must be able to move in both directions at the
same time.  Thus, the total crossbar area is 2 *
\piEightCatPrepTotalHeight\ + 2 * \piEightDecodeTotalHeight\ + 2 *
\piEightDecodeTotalHeight\ = \piEightTotalCrossbarArea\ macroblocks,
and the total functional unit area is \piEightCatPrepTotalArea\ +
\piEightTransXSZTotalArea\ + \piEightDecodeTotalArea\ +
\piEightHMTransZTotalArea\ = \piEightTotalComputeArea\ macroblocks,
resulting in a total area of \piEightTotalArea\ macroblocks.
%\begin{eqnarray} Total Area = 
%\piEightTotalCrossbarArea\ + \piEightTotalComputeArea\ = \piEightTotalArea\ macroblocks.
%\end{eqnarray}
Note that this is only
%We note, however, that this is merely 
the area for turning an encoded
zero into an encoded $\pi/8$.  This factory needs to be supplied by
zero ancilla factories in order to function, which we account for in
%our final analysis in 
Section~\ref{sec:arch}.

The bottleneck of this ancilla factory is the Cat State Prepare stage.
Each seven-qubit cat state produced by this stage results in one
encoded $\pi/8$ ancilla produced by the factory, so the throughput of
the factory is equal to the throughput of this stage:
\piEightTotalBWLogical\ encoded $\pi$/8 ancillae / ms.

As mentioned in Section \ref{sec:pi2k}, we build up smaller angle
$\pi/2^k$ rotations from combinations of $\pi/8$ and H gates instead of building
ancilla factories for them.  It is worthwhile to note that if physical
gates with adequate precision are available, the critical path for the
data can be decreased further.  From Figure \ref{fig:pi_n_anc_prep} we
see that the critical path for the data through such a factory would
on average consist of $\sum_{i=0}^{k-2}1/2^k$ CX gates and one fewer X gates.

%Abbreviated table
\begin{table}
%\begin{minipage}{0.49\hsize}
\small
\begin{center}
\begin{tabular}{|l|c|c|c|}
\hline
 & Unit & Total & Total \\
Stage & Count & Height & Area \\
\hline \hline
Cat State Prepare & \piEightCatPrepCount & \piEightCatPrepTotalHeight & \piEightCatPrepTotalArea \\
Transversal CX/CS/CZ/$\pi/8$ & \piEightTransXSZCount & \piEightTransXSZTotalHeight & \piEightTransXSZTotalArea \\
Decode (plus Store) & \piEightDecodeCount & \piEightDecodeTotalHeight & \piEightDecodeTotalArea \\
H/M/Transversal Z & \piEightHMTransZCount & \piEightHMTransZTotalHeight & \piEightHMTransZTotalArea \\
\hline
\end{tabular}
\end{center}
\vspace{-0.2in}
\caption{The functional unit counts and characteristics for each stage
of our final $\pi/8$ ancilla factory.}
\label{table:pi8_func_unit_counts}
%\end{minipage}
\hfill
\end{table}

\section{Architectural Trade-offs}\label{sec:arch}

%% {\bf
%% \begin{itemize}
%% \item FIG: microarch: data, data interconnect, anc fac, anc distribution NW
%% \item analysis: pipelined with teleport vs. non-pipe with channels vs. ???
%% \item simulation results to verify realism
%% \end{itemize}
%% }

We now bring our analyses together to draw quantitative conclusions
about running a quantum circuit at the speed of data and to compare
against proposed architectures from prior work.  Following that, we
present a more qualitative discussion of some conclusions we've drawn
from this work.

\subsection{Matching Production to Need}

\begin{figure*}
\begin{center}
\epsfig{file=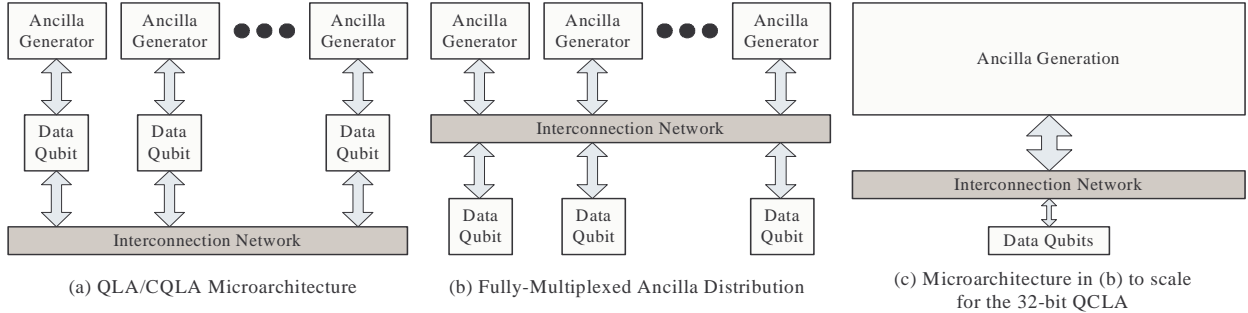,width=\hsize}
\end{center}
\vspace{-0.25in}
\caption{A quantum layout microarchitecture may be considered to
consist of three components: generators of encoded ancillae, data
qubit computation regions and interconnect.  (a) The (C)QLA
microarchitecture dedicates an ancilla generation unit to each data
qubit.  (b) Our general microarchitecture redirects encoded ancillae
to wherever they're needed on the chip, thus avoiding idle generators.
(c) In order to run at the speed of data, the ancilla generation
portion of the chip needs far more hardware than the data regions, as
shown in Table~\ref{table:areaForAvgBW}.}
\label{fig:3_microarchs}
\end{figure*}

We divide the microarchitecture of a quantum layout into three
components: 1) hardware resources for generation of encoded ancillae;
2) hardware resources for data operations, including operations
involving data ancillae and the data/ancilla interaction portion of a
QEC step; and 3) an interconnection network for moving around both
encoded data and ancillae.  Figure~\ref{fig:3_microarchs}a shows the
(C)QLA microarchitecture \cite{metodi2005qla,thaker2006qmh} using
these components, with each data qubit (whether in a compute region or
memory) having an associated ancilla generation unit for QEC.
Figure~\ref{fig:3_microarchs}b shows an ancilla factory-based
microarchitecture wherein encoded ancillae are being generated across
the chip and distributed to data as need dictates.
%This of course puts a greater strain on the interconnect network,
%which we shall return to in Section~\ref{???}.  For now, we assume
%that the interconnect can successfully route qubits around in a timely
%manner.

\begin{table*}
\small
\begin{center}
\begin{tabular}{|l|c|c|c|c|}
\hline
 & Encoded Ancilla & Data Area & QEC Ancilla Factories & $\pi/8$ Ancilla Factories \\
Quantum Circuit & Bandwidth For QEC & (\% of total) & Area (\% of total) & Area (\% of total) \\
\hline \hline
32-Bit QRCA & \qrcaAvgBWNeedForQECLogical & \qrcaDataArea\ (\qrcaAvgAreaPercentData\%) & \qrcaAvgZeroAncFacArea\ (\qrcaAvgAreaPercentZeroAncFac\%) & \qrcaAvgPiEightAncFacArea\ (\qrcaAvgAreaPercentPiEightAncFac\%) \\
32-Bit QCLA & \qclaAvgBWNeedForQECLogical & \qclaDataArea\ (\qclaAvgAreaPercentData\%) & \qclaAvgZeroAncFacArea\ (\qclaAvgAreaPercentZeroAncFac\%) & \qclaAvgPiEightAncFacArea\ (\qclaAvgAreaPercentPiEightAncFac\%) \\
32-Bit QFT & \qftAvgBWNeedForQECLogical & \qftDataArea\ (\qftAvgAreaPercentData\%) & \qftAvgZeroAncFacArea\ (\qftAvgAreaPercentZeroAncFac\%) & \qftAvgPiEightAncFacArea\ (\qftAvgAreaPercentPiEightAncFac\%) \\
\hline
\end{tabular}
\end{center}
\vspace{-0.2in}
\caption{Area breakdown to generate encoded ancillae at the QEC
bandwidths shown in Table~\ref{table:averageBWNeededForQEC}.  The
$\pi/8$ ancilla bandwidth is computed to match.  The last column
includes area for both $\pi/8$ encoding and the zero ancilla factories
supplying these encoders.}
\label{table:areaForAvgBW}
\end{table*}

Table~\ref{table:areaForAvgBW} gives the relative areas of two of the
three components of the microarchitecture in
Figure~\ref{fig:3_microarchs}b when running our benchmarks at (or
near) the speed of data under our ion trap assumptions.  We depict our
microarchitectural components to scale for the 32-bit QCLA in
Figure~\ref{fig:3_microarchs}c.  The encoded zero ancilla bandwidth
for error correction is the average bandwidth required for each
circuit (Table~\ref{table:averageBWNeededForQEC}).  A corresponding
encoded $\pi/8$ ancilla bandwidth is computed (but not shown in the
table) to run the circuit at that speed.  Column 4 includes only those
zero ancilla factories producing for QEC.  Column 5 includes both
$\pi/8$ encoding factories and sufficient encoded zero ancilla
factories to supply the $\pi/8$ encoding factories.

We see that even the most serial of the benchmarks, the Quantum
Ripple-Carry Adder, requires a substantial portion of the chip
(two-thirds) dedicated to encoded ancilla generation in order to take
this generation off the execution's critical path, while the more
parallel QCLA requires more than 90\%.

\subsection{Latency/Area Evaluation}

The proposals for both QLA and CQLA specify space for only serial
production of ancillae at each encoded data qubit location.  We
generalize this to GQLA and GCQLA in which we replicate the ancilla
area at each data qubit to allow parallel production of ancillae.
CQLA has additional flexibility in that different numbers of data
units can be present in the compute cache.  We wish to quantify the
efficiency of ancilla production in each microarchitecture by studying
area needed for a given execution time.

\paragraph{Methodology:}
Using dataflow graphs of our benchmarks and the estimates in
Tables~\ref{table:func_units}-\ref{table:pi8_func_unit_counts}, we
implemented an event-based simulation of ancilla factory production
and data qubit gate consumption.
%% Since Qalypso has consolidated ancilla production, our simulation
%% allows data consumption to proceed in dataflow order at the rate the
%% aggregate bandwidth of the ancilla factories (Tables
%% \ref{table:func_unit_counts} \& \ref{table:pi8_func_unit_counts}) can
%% accommodate.
Simulation of the QLA \cite{metodi2005qla} microarchitecture assumes
that each data qubit in the computation has a dedicated cell with
ancilla production.  Data qubits are always moved back to their home
base to do the error correction after each encoded gate. We simulate
dataflow execution taking into account latency of the ancilla
production and encoded gate execution, using latencies from Tables
\ref{table:func_units} and \ref{table:pi8_func_units}.
%We use the same operation latencies as our layout even though QLA uses
%a different, non-pipelined data cell.  We consider this optimistic in
%favor of QLA because of the pipelining benefits outlined in Section
%\ref{sec:pipe_fac}.

CQLA \cite{thaker2006qmh} optimizes the QLA design by adding a cache
of data qubits that are in the current working set.  To simulate this,
we added tracking of which qubits are in the ``compute cache'' and
account for cache miss and write-back latencies.  This was the most
complicated simulation and has an implementation similar to that of
\emph{sim-cache} in SimpleScalar \cite{burger1996efm}.  We used the
same basic ancilla production and data gate latencies as for QLA.

\begin{figure*}
\begin{minipage}{0.33\hsize}
\begin{center}
\epsfig{file=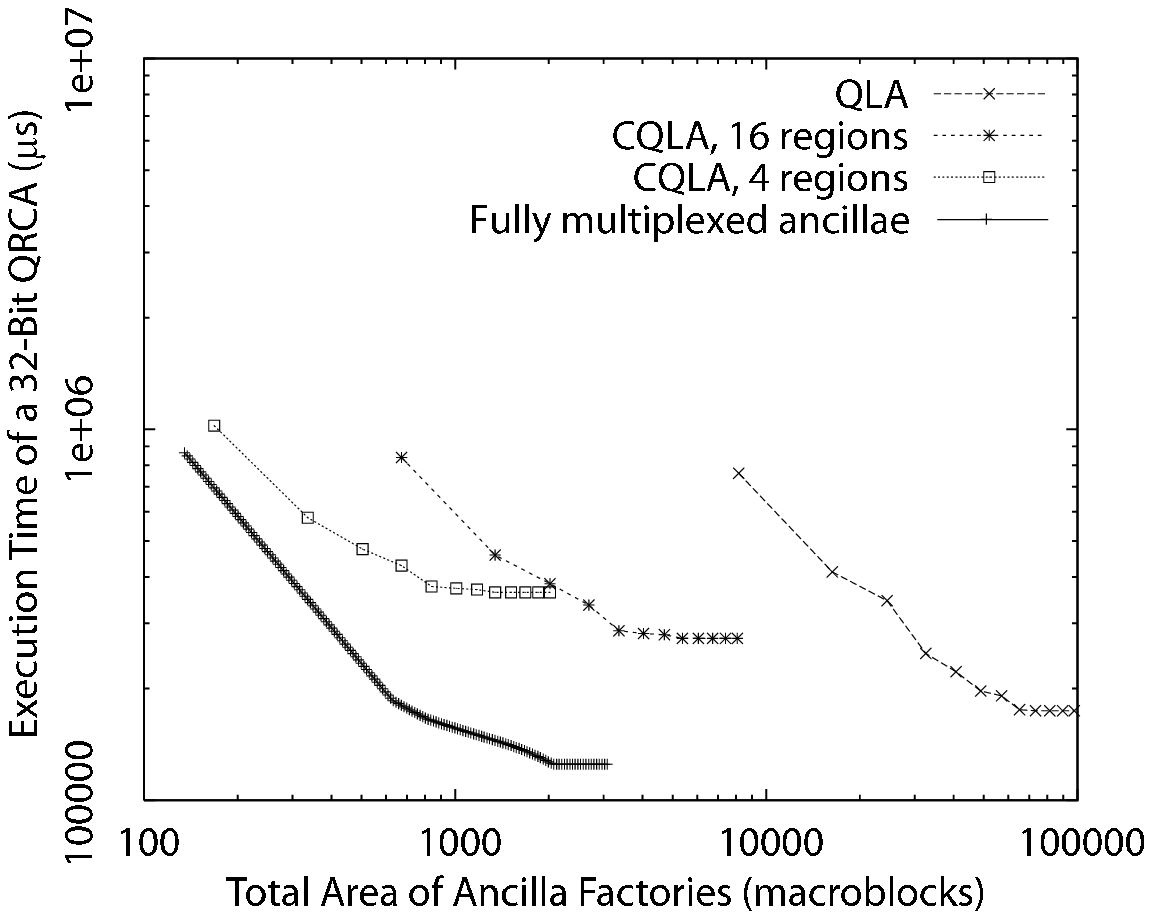,width=\hsize}
\end{center}
\vspace{-0.2in}
%\caption{(a)}
%\caption{Execution time of the 32-bit QRCA circuit for varying area
%dedicated to ancilla factories (both zero and $\pi/8$ factories).
%Data qubit area is constant at \qrcaDataArea\ macroblocks.}
\label{fig:qrcaTimeArea}
\end{minipage}%
%\end{figure}
\hfill%
%\begin{figure}
\begin{minipage}{0.33\hsize}
\begin{center}
\epsfig{file=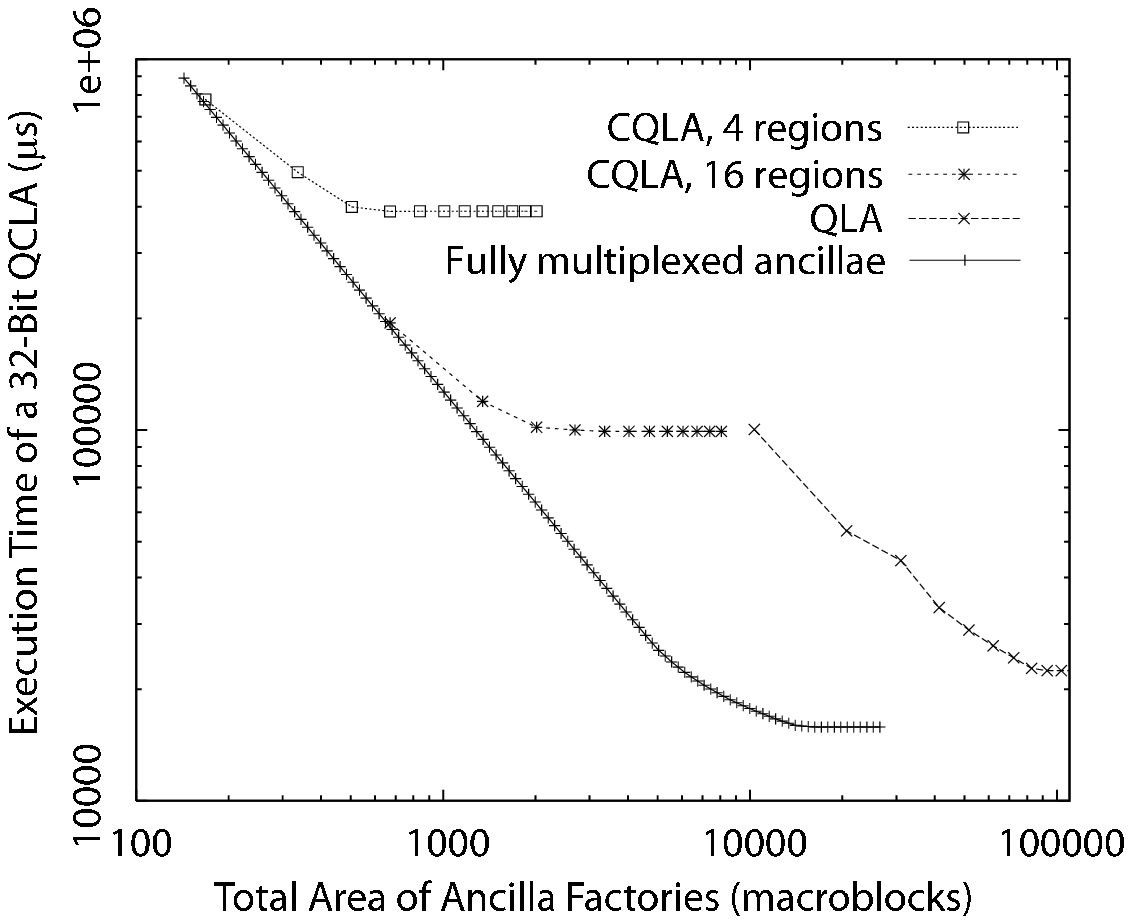,width=\hsize}
\end{center}
\vspace{-0.2in}
%\caption{(b)}
%\caption{Execution time of the 32-bit QCLA circuit for varying area
%dedicated to ancilla factories (both zero and $\pi/8$ factories).
%Data qubit area is constant at \qclaDataArea\ macroblocks.}
\label{fig:qclaTimeArea}
\end{minipage}%
%\end{figure}
\hfill%
%\begin{figure}
\begin{minipage}{0.33\hsize}
\begin{center}
\epsfig{file=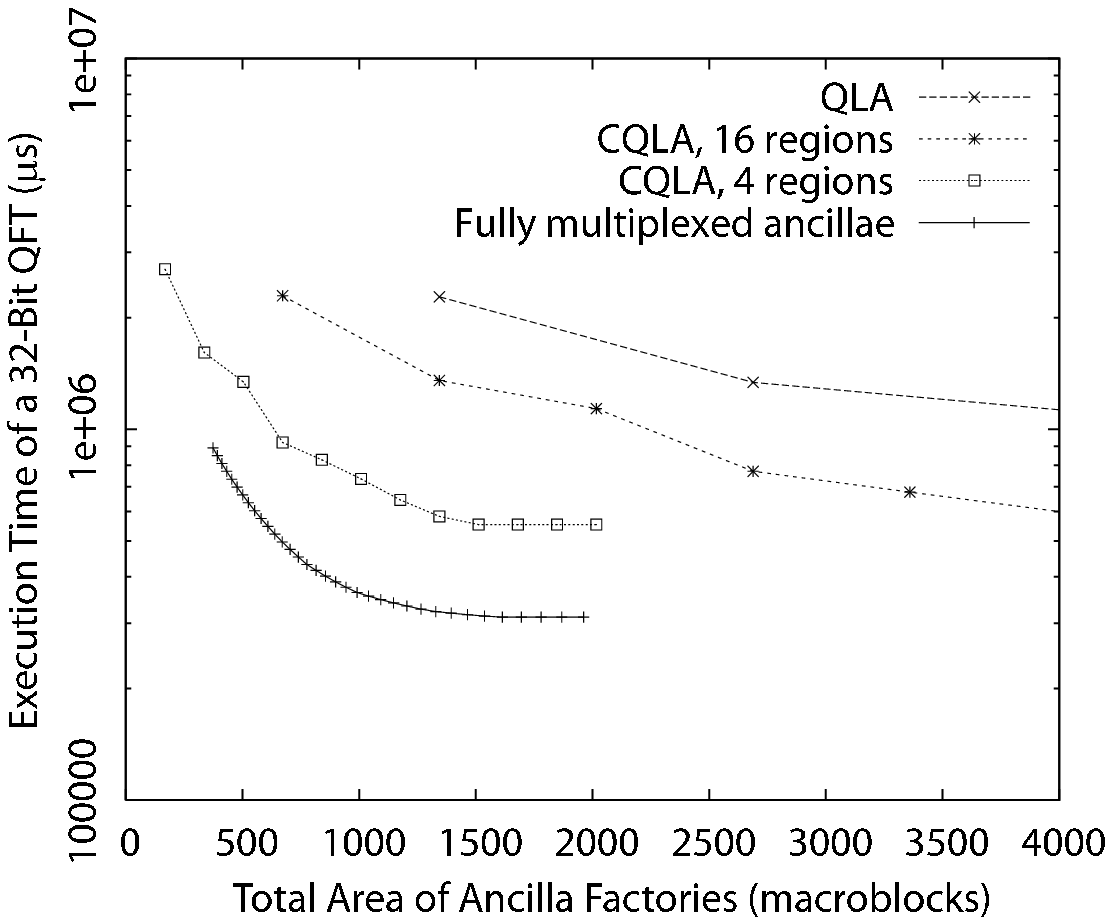,width=\hsize}
\end{center}
\vspace{-0.2in}
%\caption{(c)}
%\caption{Execution time of the 32-bit QFT circuit for varying area
%dedicated to ancilla factories (both zero and $\pi/8$ factories).
%Data qubit area is constant at \qftDataArea\ macroblocks.}
\label{fig:qftTimeArea}
\end{minipage}%
\caption{Execution time as a function of total area of encoded ancilla
factories.  (Left) 32-bit QRCA, Data qubit area = \qrcaDataArea\
macroblocks; (Middle) 32-bit QCLA, Data qubit area = \qclaDataArea\
macroblocks; (Right) 32-bit QFT, Data qubit area = \qftDataArea\
macroblocks.}
\label{fig:allTimeArea}
\end{figure*}

\paragraph{Results:}
Figure~\ref{fig:allTimeArea} shows overall circuit execution time as a
function of total area dedicated to ancilla factories (of both types)
for the different microarchitectures being tested for QRCA (left),
QCLA (middle) and QFT (right).  Total data qubit area is given in the
caption for each.

We notice that CQLA takes about half an order to an order of magnitude
longer to execute than Fully-Multiplexed Ancilla Distribution.  This
is due to the incurrence of cache misses in CQLA, whereas
Fully-Multiplexed always distributes encoded ancillae to data
when necessary.  CQLA also plateaus half an order to an order of
magnitude higher than Fully-Multiplexed since, even with very fast
encoded ancilla production, cached misses are still incurred to bring
ancillae to data.

QLA requires two orders of magnitude more area for ancilla production
to match execution time with Fully-Multiplexed, which is logical since
many ancilla generators are idle much of the time in QLA when they
could be used to feed nearby data need.  On the other hand, QLA
eventually plateaus at a similar execution time as Fully-Multiplexed,
which makes sense since it has no concept of cache misses.  QLA simply
needs very high encoded ancilla production at each data qubit in order
to run at the speed of data.

\subsection{Qalypso: Microarchitectural Implications of Pipelined Ancilla Factories}
The simple encoded zero ancilla factory in
Figure~\ref{fig:anc_fac_line} has an area of 90 macroblocks and a
throughput of \ancFacSimpleBWLogical\ encoded ancillae per
millisecond.  The pipelined encoded zero ancilla factory designed in
Section~\ref{sec:pipe_fac} has an area of \totalAncFacArea\
macroblocks and a throughput of \totalAncFacBWLogical\ encoded\
ancillae / ms.  They produce virtually the same encoded zero ancilla
bandwidth per unit area, thus seemingly negating some of the benefits
of pipelining\footnote{This is a result of the facts that the
technology is inherently synchronous and that individual gate
locations are multi-purpose.}.

%This is in fact true for two reasons.  First, due to the precise
%electrode and laser pulse sequences needed to implement movement and
%gates, respectively, our layout is by definition synchronous.
%Pipelining usually adds synchrony to an asynchronous circuit, so we're
%synchronizing an already synchronous circuit.  Second, each stage of a
%classical pipeline is usually capable of performing one subtask of the
%overall task.  In our case, each stage is by default capable of
%performing the entire task, thus we have a lack of specialization by
%stage, resulting in the non-pipelined layout being equal in size to a
%single stage of the pipeline.

%% , reaffirming the observation made in
%% Section~\ref{sec:pipe_fac} that the synchronous nature and lack of
%% gate specialization in ion trap quantum computers negates the standard
%% benefits of pipelining.

\begin{figure*}
\begin{center}
\epsfig{file=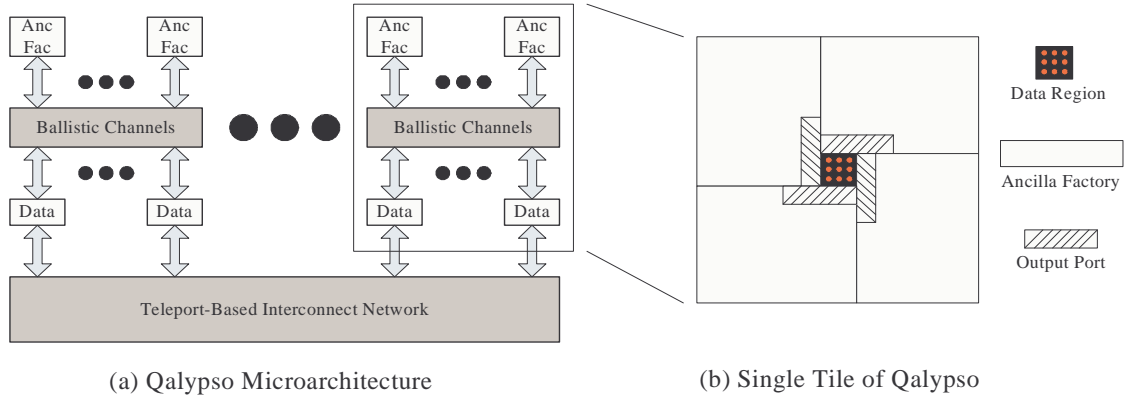,width=0.9\hsize}
\end{center}
\vspace{-0.2in}
\caption{(a) Qalypso: our proposed microarchitecture.  (b) A single
tile consists of a dense data region surrounded by ancilla factories
funneling encoded ancillae as need arises.  Ancilla distribution is
fully multiplexed within each tile, with factory output ports placed
physically close to the data region.}
\label{fig:arch_final}
\end{figure*}

Nonetheless, we conclude that pipelined ancilla factories provide
significant benefit in having concentrated input and output ``ports.''
We propose Qalypso, a tiled microarchitecture shown in
Figure~\ref{fig:arch_final}a using the tile shown in
Figure~\ref{fig:arch_final}b, with ballistic movement being used
within a tile and teleportation of data between
tiles~\cite{isailovic2006ins}.  The central data region consists of a
dense packing of encoded data qubits and channels for local ballistic
movement.
%and enough channels for ballistic movement between them.
The ancilla factories each have an output port physically near the
data region so encoded ancillae do not have far to travel.  This is
beneficial both in reducing aggregate movement error on encoded
ancillae and in avoiding congestion problems from having encoded
ancillae generated uniformly throughout an ancilla factory.
%% The output port approach would be alleviated by a careful ancilla
%% factory layout which takes advantage of the fact that only one-third
%% of physical qubits entering the last stage of a zero ancilla factory
%% are actually interacted with data, but that is beyond the scope of
%% this paper.
Meanwhile, since the limiting factor on move speed in ion traps is
state decoherence rather than control of the electrodes, stateless
qubits may be recycled to factory input ports much more quickly,
allowing the input ports to be far from the data.

This architecture differs from (C)QLA in two significant respects.
First, our data regions consist of data alone.  In CQLA, the compute
regions consist of both data and ancilla generation units, meaning
that data are physically quite a bit further apart even within one
compute region and generally require teleportation for movement.  Even
if QEC were performed as part of teleportation~\cite{bennett1996mse}, this requires twice as
many encoded ancillae as a straightforward QEC step.  Thus, we suggest
that our data regions be made as large as possible to allow data
qubits to reach each other using ballistic movement instead of
teleportation as much as possible.  Though ballistic movement is
somewhat error prone, the area of a data region consisting of
nothing but encoded data qubits is still quite small, so teleportation
is only necessary between data regions.

Second, ancilla factories surrounding a data region in our design are
shared by all data qubits within that region.  In
Figure~\ref{fig:3_microarchs}a, which represents the (C)QLA
microarchitecture, each ancilla generator is dedicated to a single
data qubit (location), so imbalances in encoded ancilla need cause
some generators to go idle while others cannot meet need.  By having a
full crossbar between generators and consumers (data qubits), as in
Figure~\ref{fig:3_microarchs}b, fresh ancillae go where they are
needed within a single data region.

The choice of data region size is still an open problem and depends on
the level of parallelism in the target application.  The determining
factors will likely be local movement congestion within the data
region and load on the inter-tile interconnect, which are shown as the
grey boxes in Figure~\ref{fig:arch_final}a.  Analyses concerning these
trade-offs will be the subject of future research.

\section{Conclusion}\label{sec:conclusion}

We find that ancilla generation bandwidth in a quantum computer is the
primary performance bottleneck, and we present a microarchitecture
that takes this bottleneck off the critical path.  We examine two
major consumers of ancillae: quantum error correction (QEC) and
non-traversal quantum gates, such as the $\pi/8$ gate for the [[7,1,3]]
CSS code. We characterize our benchmarks to find bandwidth needs
ranging from 30 to 300 encoded zero ancillae / ms and ranging from 7
to 60 encoded $\pi/8$ ancillae / ms for ion trap quantum computers.

Our resulting microarchitecture, Qalypso, is optimized for ancilla
generation and distribution, featuring dense data-only regions fed by
nearby ancilla factories.  We present layouts for these ancilla
factories and show that ancilla generation takes the majority of the
chip area even in the most serial of our circuits, the ripple-carry
adder.  As an interesting aside, we find that pipelining does not have
the same beneficial impact on throughput as in classical circuits but
does provide an important structural benefit: it can produce high
bandwidth ancillae directed at a single output port.  Qalypso can
produce circuits of similar speed to previous architectures with
greatly reduced resources or alternatively can produce circuits of
much greater speed than previous architectures for similar area.

{\singlespace\small
\bibliographystyle{unsrt}
\bibliography{main}
%\bibliogrpahy{main2}
}
\end{document}